\documentclass[fleqn,10.5pt]{wlscirep}

\usepackage{graphicx}
\usepackage[space]{grffile}
\usepackage{latexsym}
\usepackage{textcomp}
\usepackage{longtable}
\usepackage{multirow,booktabs}
\usepackage{amsfonts,amsmath,amssymb}
\usepackage{url}
\usepackage{hyperref}
\hypersetup{colorlinks=false,pdfborder={0 0 0}}
\usepackage[utf8]{inputenc}
\usepackage[english]{babel}
\usepackage{setspace}
\usepackage{pdfpages}

 \newcommand{\beginsupplement}{%
        \setcounter{table}{0}
        \renewcommand{\thetable}{S\arabic{table}}%
        \setcounter{figure}{0}
        \renewcommand{\thefigure}{S\arabic{figure}}%
     } 

\title{Dynamic Balance of Excitation and Inhibition  in Human and Monkey Neocortex}

\author[1,2,*]{Nima Dehghani}
\author[3]{Adrien Peyrache} 
\author[4]{Bartosz Telenczuk} 
\author[5]{Michel Le Van Quyen} 
\author[6]{Eric Halgren} 
\author[7]{Sydney S. Cash}
\author[8]{Nicholas G. Hatsopoulos}  
\author[4]{Alain Destexhe} 
  
\affil[1]{Wyss Institute for Biologically-Inspired Engineering, Harvard University, Boston, MA, USA}
\affil[2]{New England Complex Systems Institute, Cambridge, MA, USA.} 
\affil[3]{NYU Neuroscience Institute and Center for Neural Sciences, New York University, NYC, NY, USA.}  
\affil[4]{Laboratory of Computational Neuroscience, Unité de Neurosciences, Information et Complexité, CNRS, Gif-Sur-Yvette, France.}  
\affil[5]{Institut du Cerveau et de la Moelle Epinière, UMRS 1127, CNRS UMR 7225, Hôpital de la Pitié-Salpêtrière, Paris, France.}  
\affil[6]{Multimodal Imaging Laboratory, Departments of Neurosciences and Radiology, University of California San Diego, La Jolla, CA, USA.}  
\affil[7]{Department of Neurology, Massachusetts General Hospital and Harvard Medical School, Boston, MA, USA.}  
\affil[8]{Department of Organismal Biology and Anatomy, Committee on Computational Neuroscience,  University of Chicago, Chicago, IL, USA.}  

\affil[*]{correspondence should be addressed to:  nima.dehghani@wyss.harvard.edu, nima@necsi.edu}

\begin{abstract}
Balance of excitation and inhibition is a fundamental feature
of in vivo network activity and is important for its computations.
However, its presence in the neocortex of higher mammals is not well
established. We investigated the dynamics of excitation
and inhibition using dense multielectrode recordings in humans
and monkeys. We found that in all states of the wake-sleep cycle,
excitatory and inhibitory ensembles are well balanced, and
co-fluctuate with slight instantaneous deviations from perfect
balance, mostly in slow-wave sleep.  Remarkably, these correlated
fluctuations are seen for many different temporal scales. The
similarity of these computational features with a network
model of self-generated balanced states suggests that such balanced
activity is essentially generated by recurrent activity in the local
network and is not due to external inputs.  Finally, we find that this
balance breaks down during seizures, where the temporal correlation of
excitatory and inhibitory populations is disrupted.  These results
show that balanced activity is a feature of normal brain activity, and
break down of the balance could be an important factor to define
pathological states.
\end{abstract}

\begin{document}
\flushbottom
\maketitle
\thispagestyle{empty}

\section*{Introduction}
It is believed that neuronal networks \textit{in vivo} function in a "balanced" regime, where excitatory and inhibitory neuron activities maintain tightly correlated levels of activity.  This balanced excitation/inhibition (E/I) was first suggested theoretically \cite{Shadlen1994,Vreeswijk1996} and later found experimentally in vitro \cite{Shu2003} and in vivo \cite{Haider2006}. It is not only considered to be a functional cornerstone in the cerebral cortex, but also has been hypothesized to play a major role in areas other than cortex \cite{Okun2010}.

Whether or not this concept of E/I balance can be extended to higher mammals, such as monkey or humans, is presently unknown. In this paper, we address this question by taking advantage of recent advances in the neural ensemble recordings
with multi-electrode systems \cite{Nicolelis2008} and the ability to separate excitatory and inhibitory cells \cite{Bartho2004,Peyrache2012} in order to characterize the dynamics of excitatory and inhibitory populations, in human recordings (temporal cortex), and monkey recordings (motor and pre-motor cortex) \cite{Takahashi2015}. The units were initially clustered based on spike shape, and in a next step, their excitatory or inhibitory character could be confirmed by their functional interactions, as determined using cross-correlograms \cite{Peyrache2012}. To the best of our knowledge, this procedure provided the first coherent separation between identified populations of excitatory and inhibitory cells in humans.  This was only possible because of the long period of the recordings (several segments of continuous 12-hour recordings for each subject).  A similar discrimination between RS and FS cells was also done for the monkey recordings using a similar electrode array (see \cite{Takahashi2015}). Together, these human and monkey recordings provide a unique data set where one can investigate the dynamics of excitation and inhibition in different brain states.  In the present paper, we characterize the dynamics of ensemble inhibition and excitation at many temporal scales, analyze their interaction in different brain states and characterize the situations when the balance breaks down.

\section*{Results}

We first show the dynamic balance between excitatory and inhibitory
cell activities in all different brain states, in human and monkey.
We then use a number of methods to quantify this balance, as well as
the deviations from balanced activity.  Finally, we show an example of
a pathological brain state where the balance breaks down.

\subsection*{Recordings from different states are suggestive of excitatory and inhibitory balance}
\begin{figure}[h!]
\begin{center}
\includegraphics[width=0.76\columnwidth]{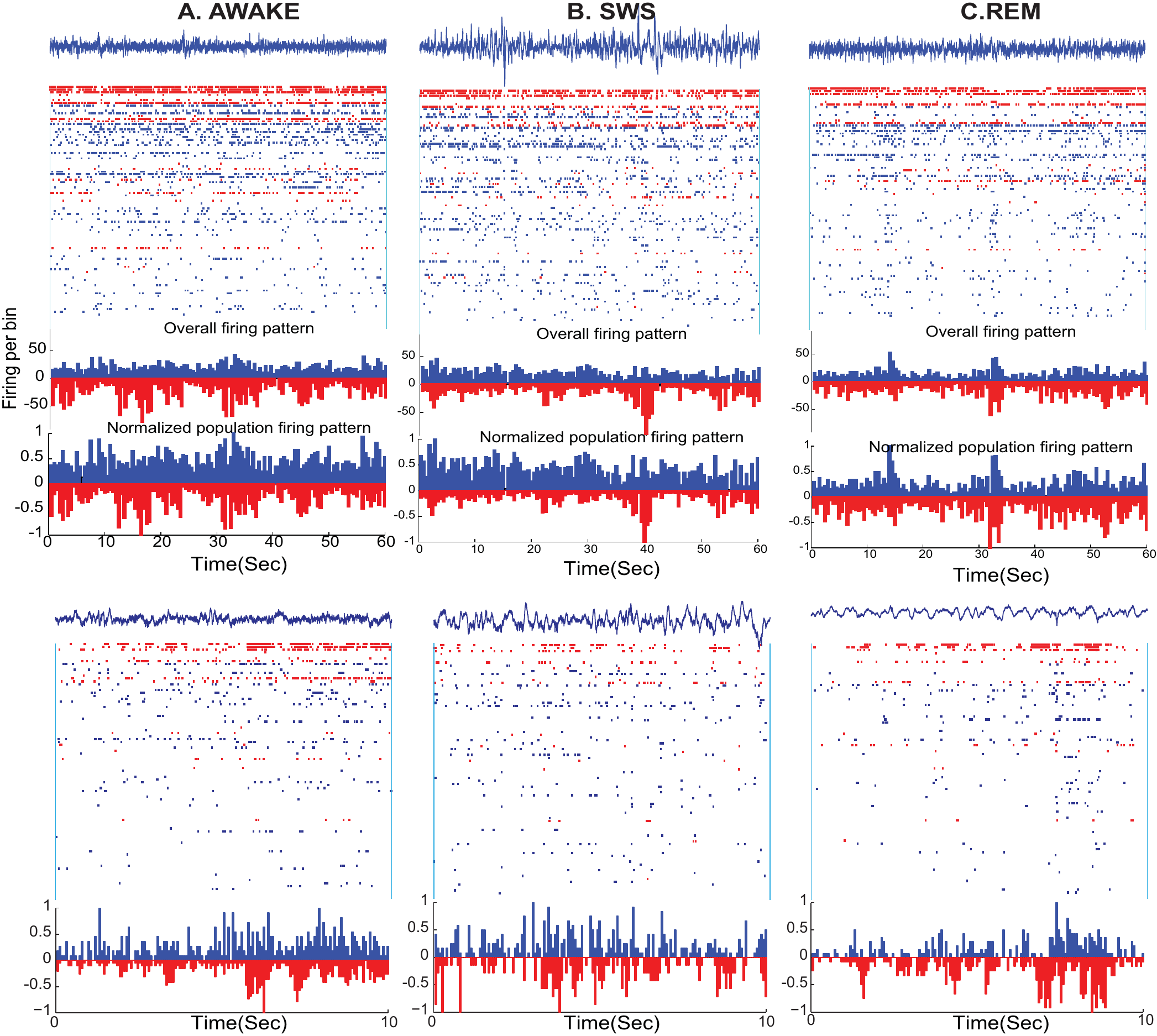}
\caption{\label{fig:StateTrace} Sample recordings for awake (A), SWS
  (B) and REM (C) in human. Top row shows 60 second windows; bottom
  row shows a 10 second window of the same state. Putative inhibitory
  neurons (FS cells) are shown in red. Putative excitatory neurons
  (RS) are depicted in blue. At the top of each panel, a sample LFP
  trace (in blue) accompanies the spiking activity. Neurons are sorted
  based on their firing rate, within the portrayed epoch, in a
  descending order. Histograms show the overall activity of RS (blue)
  and FS (red) cells. In the normalized histogram, overall activity of
  each population is normalized to the maximum of firing rate (of the
  corresponding FS or RS population) in the shown example. Zero lag
  correlation values between the ensemble E and I are respectively: 0.726,
  0.47 and 0.503.%
}
\end{center}
\end{figure}

Figure \ref{fig:StateTrace} shows local field potential (LFP) and unit recordings in human during different episodes of wakefulness, slow-wave sleep (SWS) and Rapid-Eye Movement (REM) sleep.  The rasters of unit activity is divided into RS (blue) and FS (red) cells (see human and monkey RS/FS cells spike waveforms in Supplementary Fig.\ref{fig:RawTrace} and supplementary Fig.\ref{fig:FSRSdetails}). We used this categorized ensemble activity to quantify the neocortical balance of excitation and inhibition.

A consistent observation for different states is that the ensemble inhibition
and excitation mirror each other (Fig.\ref{fig:StateTrace} in humans;
see also Supplementary Fig.\ref{fig:MonkeyStateTrace} for monkey).
One can see from the overall firing patterns (bottom), that in general
an increase or decrease of the excitatory population is mirrored by
similar dynamics among inhibitory cells, sometimes with a slight instantaneous deviation from balance (see below for quantification). Additionally, our analyses show that cells do not have a constant firing rate ratio throughout the recordings (supplementary Fig.\ref{fig:firingVariability}). Thus the cells that show high firing rate at a given time period, do not necessarily have a higher firing rate. Variability of firing rate is a feature of both E and I subgroups. As the effect of high firing rate cancels out due to this variability, the estimates of the ensemble balance is not affected by a highly dominant group of cells. 

\begin{figure}[h!]
\begin{center}
\includegraphics[width=0.4\columnwidth]{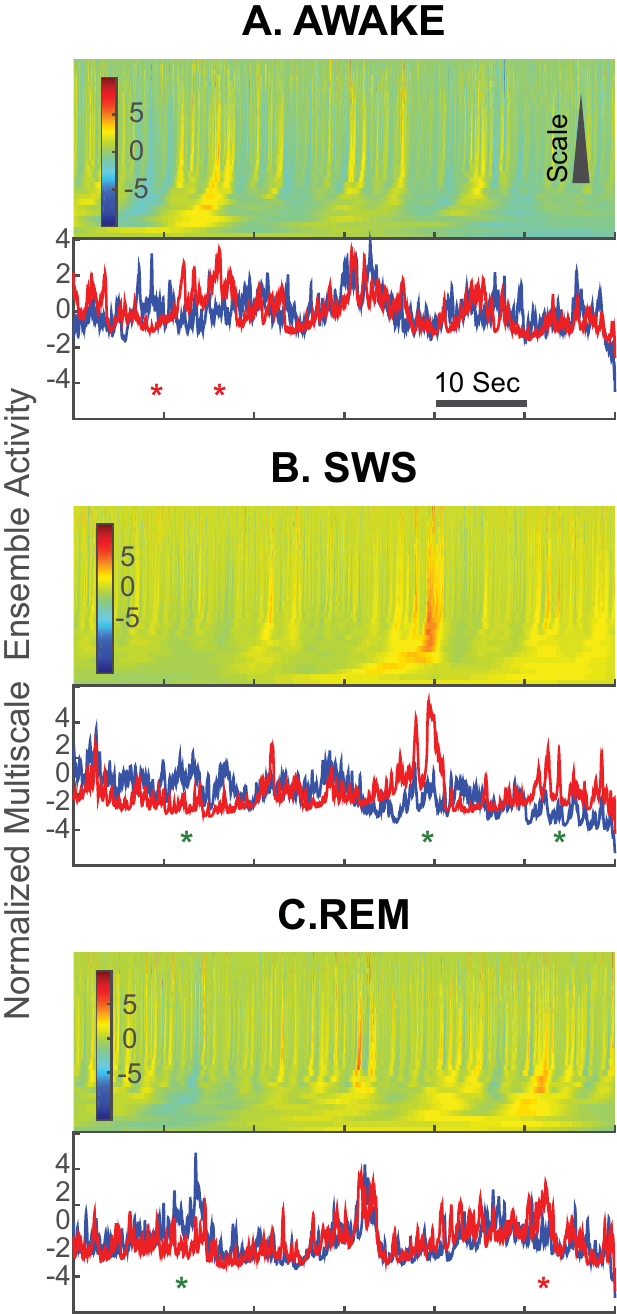}
\caption{\label{fig:rawMultiscaleTrace} Balanced excitation and
  inhibition in sample 60 seconds recordings in human (The examples
  are from Fig.\ref{fig:StateTrace}, panels A, B and C).  Heatmaps at
  the top: Each row in the heatmap shows the normalized (Z-score)
  difference of ensemble excitation and inhibition for a given scale.
  The temporal scales are defined as time bins of
    different duration, increasing from the top to bottom (from
  fine-grain to coarse-grain). For details on coarse-graining, see methods. The color saturation towards red
  signifies instantaneous dominance of inhibition. Blue saturation
  shows instantaneous dominance of excitation, while green shows tight
  match between normalized ensemble excitation and inhibition. Bottom
  traces show the Z-scored addition of normalized excitatory (blue) and
  inhibitory (red) ensembles across the scales. Red stars show where
  the two interacting ensembles do not show similar trends in their
  multiscale fluctuations. Green stars show times when ensemble
  excitation and inhibition follow each other at certain scales but
  not all. Any of these states leads to a multiscale deviation from
  perfect balance. Such deviations are more pronounced for the SWS
  (see also Fig.\ref{fig:BalanceDeviation}).%
}
\end{center}
\end{figure}

A further look at the example recordings in Fig.\ref{fig:StateTrace}
shows that most of the time, the two interacting ensembles
follow the same trend at multiple scales and that deviations from
perfect balance seem more pronounced for the SWS.
Additionally, it is noticeable that sometimes the two ensembles follow
each other at certain scales but not all (Fig.\ref{fig:rawMultiscaleTrace},
bottom traces, representing the Z-scored addition of normalized
excitatory (blue) and inhibitory (red) ensembles across the scales.).
Similar patterns are observable in examples from monkey recordings
(see Fig.\ref{fig:monkeyMultiscaleTrace}). To test whether the analyses of those neurons that demonstrate very typical features of each (FS or RS) group, we chose the units only from the 30\% of the two ends of the classification spectrum (as shown in Fig.\ref{fig:FSRSsubsample}A.).  We then re-calculated the multiscale balance of E and I for the sub-set. At any given time t, we calculated the sum of the normalized difference of ensemble E and I across multiple scales. These values were turned into a histogram (as shown in Fig.\ref{fig:FSRSsubsample}B) to evaluate the distribution of dominance of E vs I. Instantaneous dominance of ensemble E is well balanced by instantaneous dominance of ensemble I throughout the whole recordings (Fig.\ref{fig:FSRSsubsample}B top row) as well as for the shown examples in FigS3 (Fig.\ref{fig:FSRSsubsample}B mid and bottom panels). This observation leads to question whether this E/I balance extends throughout all brain states, or if there are periods where the balance breaks down? In what follows, we aim to decipher such possible relations between excitation and inhibition at different temporal scales and provide a quantitative study of the dynamic aspects of E/I balance.

\subsection*{Dynamic aspects of balance}
An important property of the balanced activity is its multiscale aspect, as illustrated in Fig.~\ref{fig:rawMultiscaleTrace}.  For each brain state represented, we calculated the difference between excitatory and inhibitory activities, represented as a function of the temporal scale considered.  The multiscale aspect of the balanced activity clearly appears as the dominance of differences close to zero, with occasional deviations occurring transiently, especially in SWS.

We also evaluated the behavior of the correlations between excitation
and inhibition across different temporal scales.  As shown in
Fig.\ref{fig:ensxcorr}, the ensemble FS and RS series showed well
correlated dynamics. This type of ensemble correlation was observed
across the multiple timescales. Further, the Monte Carlo randomization
(four different types of randomization were implemented) showed that
such correlation can not be due to aggregation of spike series into
ensembles (For details of ensemble cross-correlogram and the
randomization, see methods).

The observed ensemble temporal interdependence, was seen in different
subjects with different number of FS and RS cells yet with the similar
relative RS/FS count ratio of 4 to 1 (Fig.\ref{fig:ensxcorr}A1-A4),
was multiscale (Fig.\ref{fig:ensxcorr}B1-B4), and was observed in all
states (Fig.\ref{fig:ensxcorr}C1-C4). The percentage of co-occurrence
of spikes (in the ensemble series) at the lag zero and the maximum
observed percentage of co-occurrence (whether that maximum was at lag 0
or not) showed a robust multiscale linear relationship. A linear fit
to the pooled values of lag zero vs maximum observed correlation,
yielded a cross-subject average of 0.9988$\pm$0.0134 for Awake,
0.9985$\pm$0.0147 for REM, 0.9985$\pm$0.0162 for light-sleep and
0.9977$\pm$0.02539 for slow-wave sleep. We wish to emphasize two key
findings: a) the maximum of the \textit{ensemble} cross correlation is
close to zero lag (note that this cross correlation is not calculated
as an average of pair-wise cross-correlograms).  Instead it represents
the linear correlation of the two ensemble series (at different
scales), on average, fire together and one cell population is not
following the fluctuations of the group by some fixed delay. This does
not necessitate the influence to be forced through a common input (see model of E/I balance below) another aspect is that as the data is
  coarse grained, the peak narrows and the higher correlation of the
  short delay shoulders (in comparison to long delays) dissipates.
  This phenomenon suggests that the instantaneous E-I relation
  (at the ensemble level) is tighter at coarse time scales.

\subsection*{Model of E/I balance}
\begin{figure}[h!]
\begin{center}
\includegraphics[width=0.5\columnwidth]{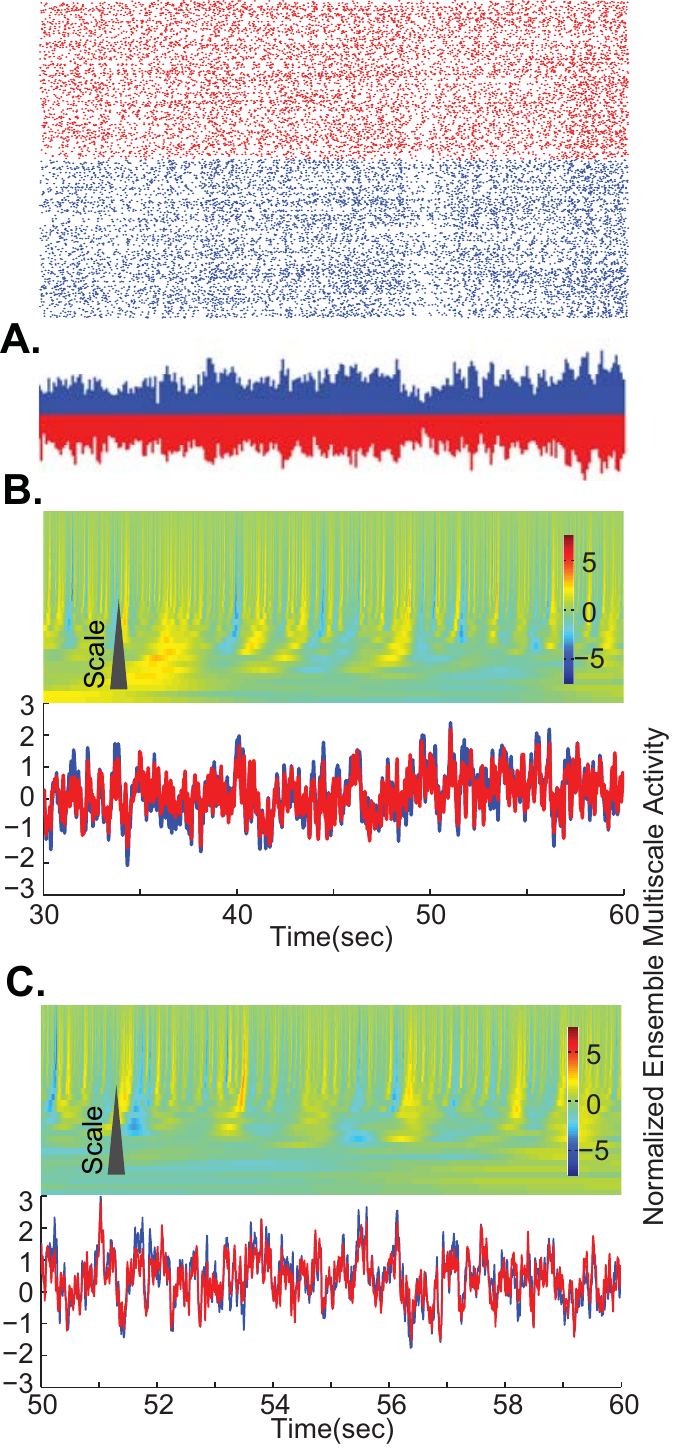}
\caption{\label{fig:COBA} Multiscale balance in a computational model
  of AI (Asynchronous Irregular) states in networks of spiking
  neurons. A. An example of raster plot and its normalized ensemble activity of COBA AI state. As in Fig.\ref{fig:rawMultiscaleTrace}, preservation of
  excitatory-inhibitory balance across scales shows mirrored activity. B. As in
  Fig.\ref{fig:rawMultiscaleTrace}, the heatmap shows the normalized
  (Z-score) difference of ensemble excitation and inhibition for
  multiple scales. Line traces show the Z-scored addition of normalized
  excitatory (blue) and inhibitory (red) ensembles across the scales.
  C. Same as B for a shorter period of time (last 10 seconds of B).
  These panels show that in general, the ensemble excitation and
  inhibition show an overall multiscale balance, even though there are
  instantaneous deviations from perfect balance.%
}
\end{center}
\end{figure}

To compare to a system with well-known and identifiable
  properties, we considered a network model of interconnected
  excitatory and inhibitory neurons displaying self-generated balanced
  activity states.  This model \cite{Vogels2005} consists of a
conductance-based (COBA) network of 4000 neurons (2000 inhibitory and
2000 excitatory neurons; see Methods for details).  In this
  model, the two population ensembles show a balanced mirrored
activity (Fig.\ref{fig:COBA}A). Further examination of the two
populations shows that the overall balance is preserved across
multiple scales (Fig.\ref{fig:COBA}B, C and
Fig.\ref{fig:COBAbalanceNew}B). Similar to the experimental data, this is
paralleled by the instantaneous deviations from perfect balance
(Fig.\ref{fig:rawMultiscaleTrace},
Fig.\ref{fig:monkeyMultiscaleTrace}) and such observations are robust
at many examined lengths of the data (see
Fig.\ref{fig:rawMultiscaleTrace}).

\begin{figure}[h!]
\begin{center}
\includegraphics[width=0.7\columnwidth]{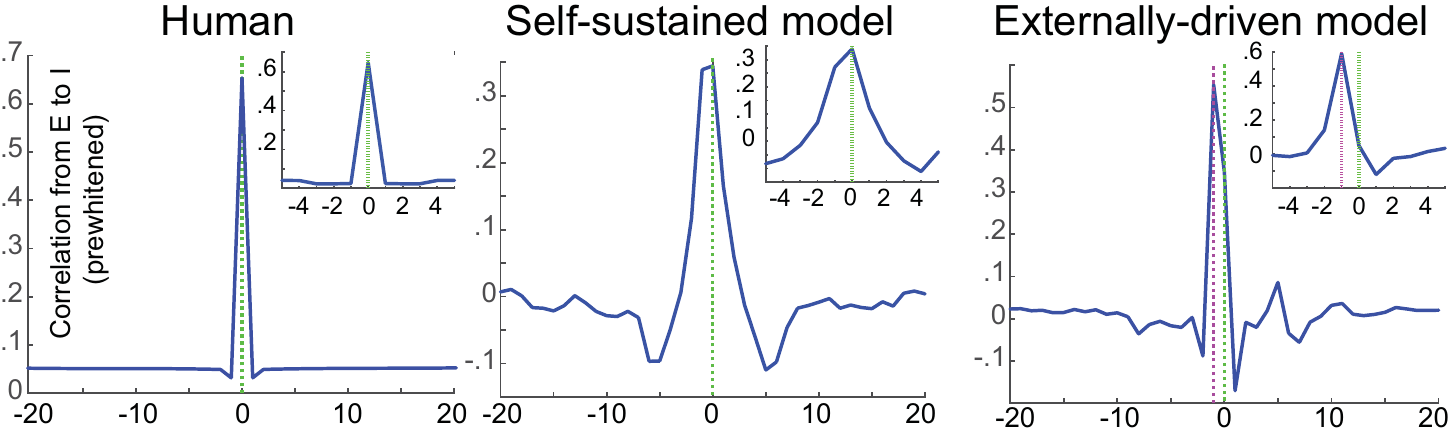}
\caption{\label{fig:CRAnew}
Correlation response analysis shows a comparative E:I impulse response for the finest time scale in Human data, self-sustained and externally-driven COBA models. In the externally-driven model, activity is mostly generated by the external inputs and stimulus is weaker on inhibitory cells. In contrast to Human data and self-sustained model, the E to I ensemble spiking correlation shows a shift from the central bin in externally-driven model. Insets show a zoomed in view.%
}
\end{center}
\end{figure}

To further probe the difference between model and data, we  computed cross-correlations between excitatory and inhibitory  populations using a similar procedure and data length as the experimental data (See Fig.\ref{fig:CRAnew},Fig.\ref{fig:COBAbalanceNew}). Using prewhitened-based correlation analysis, we note that human and self-sustained COBA model show maximal correlation at central lag. However, when activity is mostly generated by the external inputs and stimulus is weaker on inhibitory cells, the correlation maxima shows a shift from the central bin (see Fig.\ref{fig:CRAnew} insets). Because these results may appear in contradiction with previous measurements of a lag between excitatory and inhibitory inputs in cortical neurons \cite{Okun2009, Okun2010, Cruikshank2007}, we investigated this issue in the model. We measured the $g_{E}$ (excitatory conductance) and $g_{I}$ (inhibitory conductance) inside 100 sample cells in the network during self-sustained activity to better portray the conductance correlation at the level of individual cell and ensemble spiking. In some cells $g_{E}$ precedes $g_{I}$, and in some $g_{I}$ precedes $g_{E}$. However, overall, $g_{E}$ precedes $g_{I}$ ($\thicksim$3ms; see supplementary Fig.\ref{fig:conductance}C). These simulations show that the precedence of excitatory over inhibitory inputs in single neurons is compatible with the fact that the excitatory and inhibitory neurons cross-correlation peaks at time zero, and thus there is no precedence of excitation over inhibition ensemble spiking. This absence of delay of inhibition suggests that the balanced activity observed in the data mostly stems from self-generated activity in the network, as in the model. In other words, this analysis suggests that the E/I balance is mostly generated by the local network through recurrent connections.

\subsection*{Quantification of the E/I balance}
\begin{figure}[h!]
\begin{center}
\includegraphics[width=0.35\columnwidth]{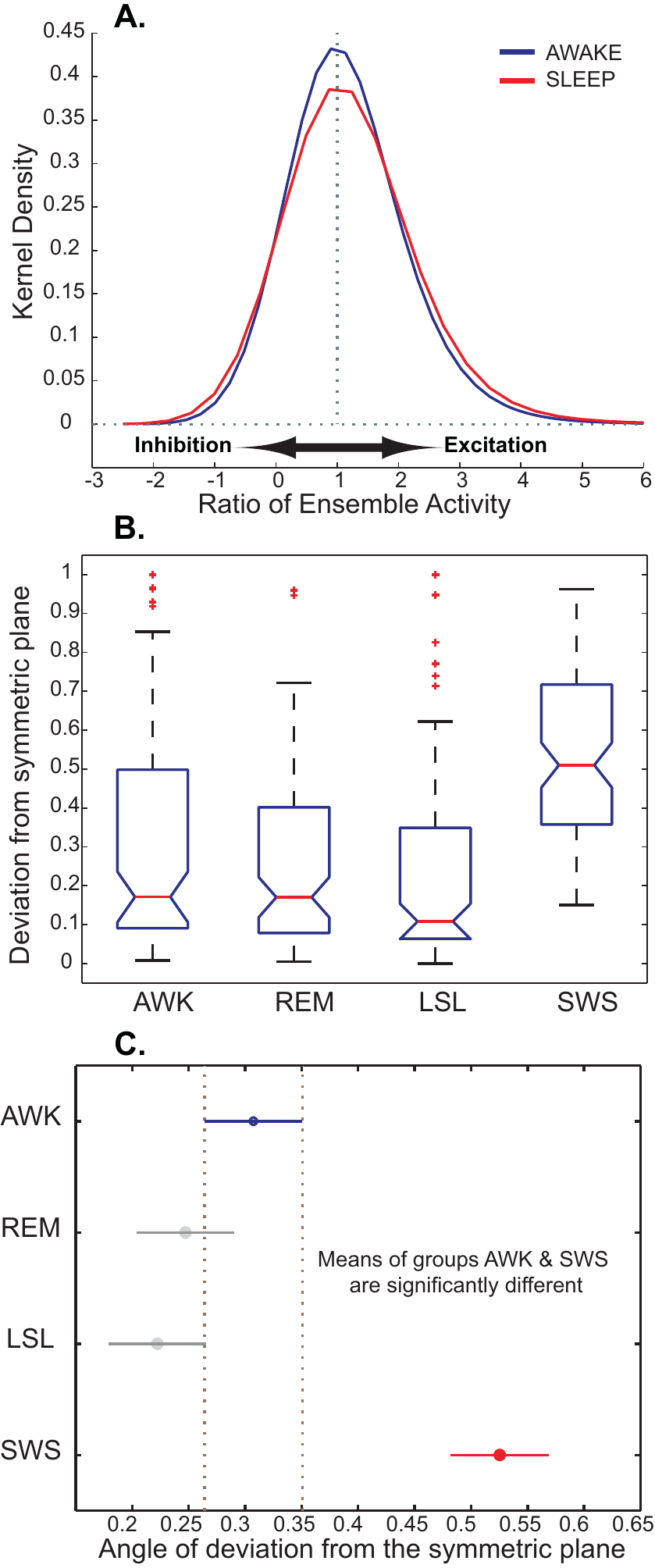}
\caption{\label{fig:BalanceDeviation} Panel A, kernel density of the ratio of E/I for the monkey awake and sleep states. Perfect balance (where the magnitude of ensemble matches) would be represented by the vertical dotted line ($=1$). Though the qualitative symmetry in each state is preserved, the kernel density estimates of sleep and awake do not match, with more kurtosis in awake and broader shoulders in sleep. A Two-sample Kolmogorov-Smirnov test on the E/I ratio at the significance level of $\alpha=0.01$, rejected ($P_{val} \ll 10^{-4}$) the null hypothesis that the data in awake and sleep are from the same continuous distribution. This is matched with the observations in humans, where the angle of deviation from the symmetric plane/axis is more pronounced during sleep rather than in awake (panel B). In the boxplot, the notch represents the median, the box boundaries show the lower and upper quartile SWS and the asterisks show the outliers. In a multiple comparison test, panel C, awake and SWS show statistically significant differences between their means ($P_{val} \ll 10^{-3}$). Note that only slow-wave sleep shows significant statistical difference with the other states at $\alpha=0.05$.%
}
\end{center}
\end{figure}

The quantification of the multiscale aspect of the balance (as in
  Fig.~\ref{fig:rawMultiscaleTrace}) shows that, although the E/I
  systems are generally balanced, there are occasional small
  deviations.  We further quantified such deviations based on
a quantification of the symmetry in joint E/I
probability space (see Supplementary Fig.\ref{fig:jointProbEI} and
details in Methods). This analysis confirmed the multiscale
  investigation of Fig.~\ref{fig:rawMultiscaleTrace}.  A
nonparametric two-sample Kolmogorov-Smirnov test, $D_{n,n'}=\max_x
|F_{AWAKE,n}(x)-F_{SWS,n'}(x)|$ where $F_{AWAKE,n}$ and $F_{SWS,n'}$
are the empirical CDFs (empirical cumulative distribution function) of
the normalized E/I ratio distributions for the two states, rejected
($P_{val} \ll 10^{-3}$) that they come from the same distribution at
the significance level of $\alpha=0.01$. This shows that the degree of
balance deviation is state-dependent.  As also shown in
Fig.\ref{fig:BalanceDeviation}, the highest degree of deviation from
perfect balance happens during SWS.

This higher degree of deviation from balance during sleep could be
attributed to the fluctuations of inhibitory/excitatory activity
during up-state and down-state \cite{Renart2010,Steriade1993,Shu2003,
  Xue2014}, as hallmarks of a bistable regime where toggling between
the two states is enforced by the mutual excitation and feedback
inhibition \cite{Wilson1972}. Transient stability of both up and
down states is the other side of the coin characterized by a rhythmic
transition between quiescent and active states \cite{Holcman2006}.
This property, leads to the observed higher degree of deviations from
absolute balance plane.

\subsection*{E/I imbalance during seizures} 

It has been speculated that the breakdown of the equilibrium between
excitation and inhibition could lead to epilepsy. The idea that the
lack of inhibition or excess of excitation can cause seizure is not a
new one \cite{Symonds1959}. This has been experimentally used to
induce or control seizures, such as for example by inducing inhibition
using optogenetics \cite{Krook-Magnuson2013, Paz2013, Tonnesen2009}.
Other optogenetic studies have related cortical E/I imbalance to other
diseases such as mood disorders as well \cite{Yizhar2011}. However,
it has been argued that such a clear-cut idea of lack of inhibition or
excess inhibition as the major frame of epileptogenesis is perhaps
misleading \cite{Engel1996}.

\begin{figure}[h!]
\begin{center}
\includegraphics[width=0.6\columnwidth]{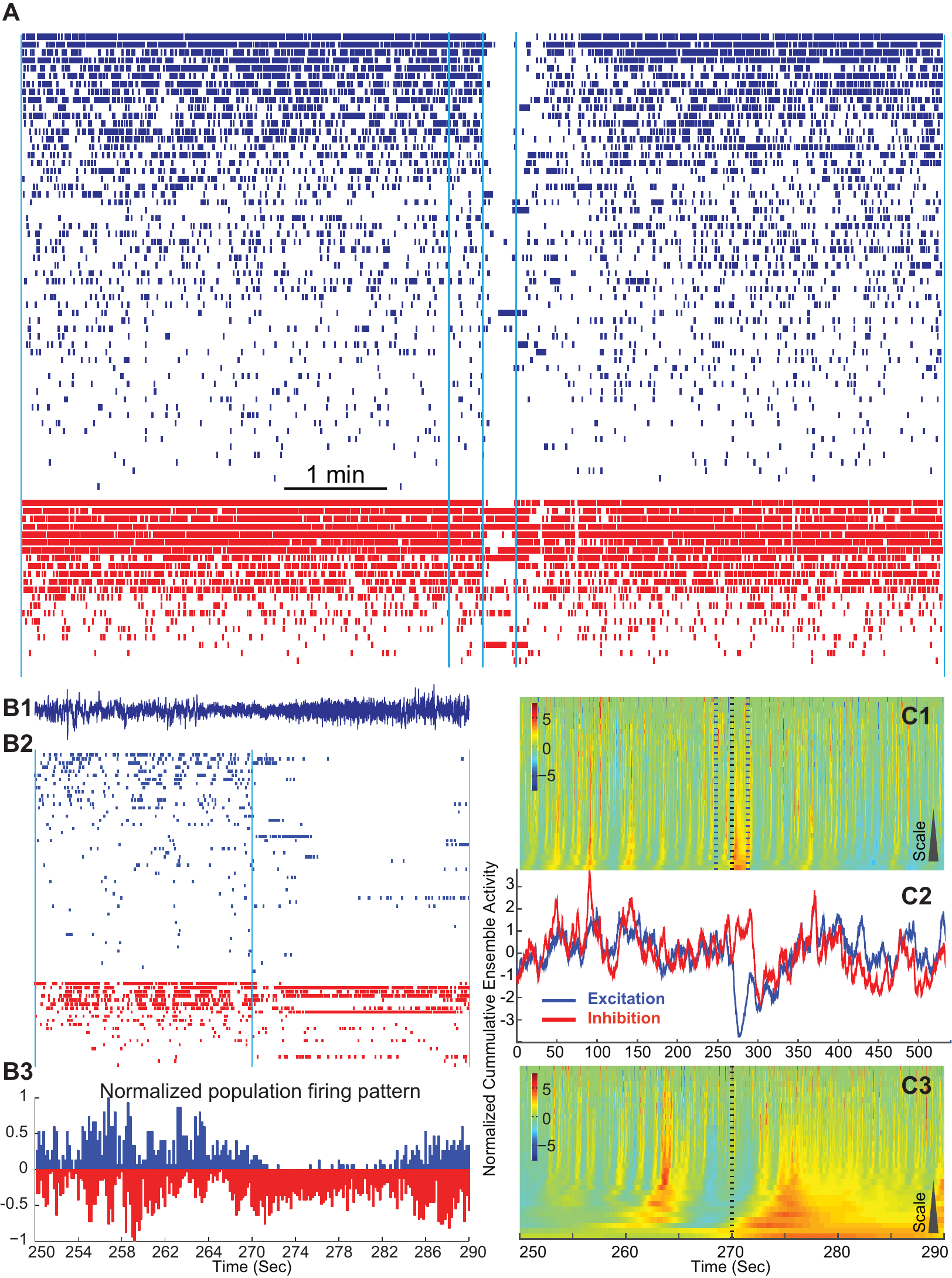}
\caption{\label{fig:SeizureMisbalance} Misbalance in an example
  seizure recording in human. Panel A shows a 9 minute recording.
  Panels in B are the zoomed in version (the middle 40 seconds) of the
  same epoch (shown with the vertical lines in A). RS cells are in
  blue and ranked based on their firing rate within this epoch. Red
  cells show FS cells and are ordered according to their class firing
  rate. B1, LFP activity in the zoomed period, corresponding to B2
  raster of FS and RS cells. B3, Normalized mirrored histogram showing
  where the misbalance occurs. C1. Heatmap of the normalized ensemble
  excitatory and inhibitory differences, corresponding to the 9 minute
  recording shown in A.  Dotted lines mark the same boundaries as in B2. C2.  Normalized cumulative ensemble activity of
  excitation vs.  inhibition during the 540 seconds epoch. C3 is the
  zoomed in version of the middle 40 seconds (corresponding to panels
  in B and the marked are by dotted lines in C1). Seizure happens
  around the mid-point and is visually distinct from the rest of the
  recording. During the seizure, a clear misbalance occurs; however it
  shows complex multiscale characteristics (see
  Fig.\ref{fig:seizureXtra} for more examples).%
}
\end{center}
\end{figure}

Here, we provide an example of a seizure recorded in one of our
patients and show how E/I balance changes in a complex fashion that is
in contrast to the simple misbalance scenario described above (see
Fig.\ref{fig:SeizureMisbalance}A,B). During the seizure some
excitatory cells and some inhibitory cells increase their firing while
some decrease or even stop firing \cite{Truccolo2011} yet an overall
imbalance persists throughout the event. The same multiscale breakdown
of the balanced excitatory-inhibitory activity was observed for all
six seizures from two human patients.  For additional examples, see
Fig.\ref{fig:seizureXtra}.

To further elaborate on the quantification of misbalance during
seizure, we tested the dynamics of the multiscale features of ensemble
excitation and inhibition throughout the seizure.  In the example
seizure (shown in Fig.\ref{fig:SeizureMisbalance}), there is a
complete break-down of the balance, were the inhibitory cells
initially dominate, which is further followed by re-emergence of
balance toward the end of this seizure episode. Heatmaps of difference
of normalized (Z-scored) ensemble excitation and inhibition,
Fig.\ref{fig:SeizureMisbalance}C1,C3 and line plots of normalized
ensemble excitation and inhibition Fig.\ref{fig:SeizureMisbalance}C2,
show that the interplay between the two populations harbors a
multiscale feature during the misbalance.

Similarly, in other examples of seizures
(Fig.\ref{fig:SeizureMultiscale}), the two ensembles follow similar
multiscale trends up to the seizure initiation (~second 270), when
suddenly the two systems become disengaged and fluctuate without any
further interdependence. Such imbalanced fluctuation is later
diminished and the two ensembles find their way to flow with the same
multiscale trend again. Though this return to the balanced trend does
not show any universal time scale. In some cases
(Fig.\ref{fig:SeizureMisbalance}C, Fig.\ref{fig:SeizureMultiscale}D),
it happens faster than others (Fig.\ref{fig:SeizureMultiscale}A), and
in other cases (Fig.\ref{fig:SeizureMultiscale}B), it may not
happen for even few minutes after the seizure has, electrographically,
ended.

Note that the particular features illustrated here, such as the
transient dominance of inhibition, may not be representative of all
types of focal seizures (and a more detailed examination of the behavior of RS and FS cells during seizures is in progress (Ahmed et al., in preparation)). As these types of ensemble recordings become
more abundant in clinical settings, in near future, it will become
possible to test the multiscale features of balance in different types
of seizures with the methods described here.

\section*{Discussion}

In this paper, we took advantage of the recent advances in the
separation of excitatory and inhibitory cells, which were confirmed by
direct cell-to-cell interaction \cite{Peyrache2012}. Our present
analysis demonstrates that the excitatory and inhibitory neural
populations are balanced in the neocortex of human and monkey, as
well as in a COBA (conductance-based) network model with AI
(asynchronous irregular) properties. This overall balance extends to
multiple temporal scales, as shown by the distributions of ensemble magnitudes
(see Fig.\ref{fig:rawMultiscaleTrace} and Fig.\ref{fig:COBA}).  We
also found that the balance extends to nearly all brain states, and
breaks down at multiple temporal scales during epileptic seizures (Fig.\ref{fig:SeizureMisbalance}). The network recovers
relatively quickly (tens of seconds) from the breakdown of
balance after the end of the seizure (Fig.~\ref{fig:SeizureMultiscale}). This pattern suggests
that the balance of excitatory and inhibitory activities is important
for normal brain function and sleep, and its breakdown is
  associated to pathological activity, although no causal link could
  be established here. 

Balance of excitation and inhibition has gone through many different
renditions \cite{Okun2009}, from simulations based on random walk
models \cite{Gerstein1964}, to opposing views of balanced synaptic
input \cite{Shadlen1994,Softky1993} and later to those relating it
to synchrony \cite{Stevens1998}, and those providing intracellular
evidence for dynamic interplay between inhibition and excitation
\cite{Rudolph2007,Monier2008}. We showed here, for the first
time, evidence for E/I balance in terms of network activity,
estimated from large ensemble of units.  

A particularly interesting finding is the absence of systematic
  phase lag between ensemble spiking of E:I populations, as shown by the zero-peaked averaged cross-correlation between RS and FS cells (see Fig.\ref{fig:ensxcorr}). This is intriguingly similar to network models where the balanced activity is self-generated (see Fig.\ref{fig:CRAnew}), suggesting that the balanced activity is mostly generated by the local network, through recurrent connections. To better link this with the issue of external inputs, we have performed additional simulations of networks of excitatory and inhibitory neurons, and compared the case of an activity completely generated `'internally'' (self-sustained asynchronous-irregular states with no external input), with the same network with 10-times reduced collateral synaptic weights, but receiving a noisy external input. When the activity was self-sustained, the cross-correlation between E and I populations (determined from spiking activity) peaks at zero (Fig.\ref{fig:CRAnew}). When network is externally-driven, ensemble spiking correlation peaks at a delayed lag (see Fig.\ref{fig:CRAnew} insets). These constitute further evidence that most of the balanced activity is generated internally by the local network. 
  
  In addition, we also found that overall, $g_{E}$ precedes $g_{I}$ in the self-sustained model (See supplementary Fig.\ref{fig:conductance}). Prior studies have reported a lag between $g_{E}$ and $g_{I}$ as well \cite{Okun2010}. A recent study also shows that this relation is state-dependent and the sign of the delay in anesthesia and awake are reverse \cite{Haider2012}. This has been ascribed to the possibility that awake activity is more driven by the thalamocortical inputs whereas in anesthesia it is driven by internal states \cite{Tetzlaff2012}. If so, the awake delay of $g_{I}$ with respect to $g_{E}$ could correspond to the feed-forward inhibition \cite{Cruikshank2007}. Like these studies, the asynchronous-irregular model presented here shows a delay between $g_{E}$ and $g_{I}$ (supplementary Fig.\ref{fig:conductance}). The absence of delay in population spiking (in both experiments and model) along with the conductance delay at the level of individual cells suggests that most of the spiking activity is self-organized by the network. Future work should investigate in detail the computational properties of such locally-balanced networks, and how their dynamics can be shifted towards unbalanced pathological activities.

\subsection*{Conclusion}
Our results suggest that excitatory and inhibitory populations 
are tightly balanced across all states of the wake-sleep cycle, in
both human and monkey.  The only times where balanced activity breaks
down is during epileptic seizures, suggesting that balanced activity
is a fundamental feature of the normal functioning brain.

\section*{Methods}
\subsection*{Recordings}

Recordings of the ensemble neural activity were obtained through the implants of multielectrode arrays (Neuroport/Utah electrodes, Blackrock Microsystems). These arrays are composed of 100 electrodes arranged in a 10x10 matrix with an inter-electrode distance of 400 microns (for more details on electrodes see \cite{Campbell1991,Jones1992}). The patients, who were implanted, suffered from intractable seizures and were under neurosurgical monitoring to localize the focus of their epileptic seizure. The electrodes tips reached layer II/III of the neocortex (for details of implants see \cite{Truccolo2011}). In the monkey, the implant was in the dorsal premotor cortex (PMd). Recordings were made during the performance of a motor task as well as during sleep (for details of implantation see \cite{Dehghani2012,Truccolo2010}). For the human studies, patients were given consent forms with detailed description of the purpose of the study and its potential risks. Approval for all human experiments involving recordings of single unit activity in patients was granted by the Institutional Review Boards of Massachusetts General Hospital / Brigham \& Women’\selectlanguage{english}s Hospital in accordance with the Declaration of Helsinki and required informed consent from each participant. For the primate experiments, all of the surgical and behavioral procedures were approved by the University of Chicago’\selectlanguage{english}s IACUC and conform to the principles outlined in the Guide for the Care and Use of Laboratory Animals (NIH publication no. 86-23, revised 1985; IACUC Approval number:  71565).

As has been described previously, the spikes of putative excitatory (Regular-Spiking., RS) neurons tend to be broader than putative inhibitory (Fast-Spiking, FS) neurons \cite{McCormick1985,Bartho2004}. The recordings were then spike-sorted and the units were categorized as either RS or FS. This categorization was based on morpho-functional characteristics of the spike-waveform and putative mono-synaptic connections (for details of such techniques see \cite{Bartho2004,Peyrache2012}). A variety of extracted features describing the shape of the average spike waveform were used, such as half-width of the positive peak, half-width of the negative peak, interval between negative and positive peaks (valley-to-peak) and the ratio of the negative to positive peak amplitude. Among the different criterion used to distinguish between cortical cell types in extracellular recordings, waveform duration is among the most reliable \cite{Bartho2004}. Some excitatory cells of motor may exhibit narrow spikes, but these are rare cases found in motor cortex \cite{Vigneswaran2011}. In addition, some  subtypes of inhibitory, non-fast-spiking interneurons show broad waveforms \cite{Fuentealba2008}. However, as these cells represent at most half of the GABAergic neuronal population \cite{Jiang2015} and that, overall, inhibitory interneurons represent about 20\% of all cortical neurons, the false positive rate of excitatory cell classification is at most 10\%. Narrow-spike neurons encompass with high confidence various types of fast-spiking (and paravalbumin-expressing) neurons such as basket cells \cite{Royer2012}. Based on these parameters, we classified the spike waveforms of all neurons into two groups using a standard K-means clustering algorithm. The procedure was repeated for each recording session separately; the neurons that were not assigned consistently to the same group were removed from further analysis.

\subsection*{Multiscale temporal rescaling}
We used 32 different time scales to remap the ensemble activity to renormalized time-series of excitation and inhibition. The scales were equally spaced in a logarithmic fashion between 1ms to 10938 ms. The logarithmic spacing was chosen for computational efficiency with respect to the number of scales, leading to a denser spread in finer time resolution. For this process, spikes of ensemble excitatory group were binned at different time-scales. As the number of excitatory and inhibitory neurons in each recording differs from one another (even though their relative size ~4/1, is close to the anatomical observations), these values were normalized by the number of the neurons in each category of cells to obtain the ensemble fraction. This condition would overcome the limitations arising from both sub-sampling (here, hundreds of neurons out of many thousands) and spatial non-uniformity of sampling (although the recording electrode is a regular grid, unit recordings are not always regularly spaced). The same process was repeated for inhibitory neurons. The results yielded ensemble fractions of inhibition and excitation at many different time scales. 

\subsection*{Ensemble cross correlation}

We first created the ensemble pool of the FS and RS cells in each
subject of the study. The two series were lined up temporally along a
common time axis. The ensemble RS and FS cells were used as the
reference and target series respectively. For a given temporal scale,
the bin length was defined according to the size of that scale as in
Fig.\ref{fig:rawMultiscaleTrace}. For each spike in the reference
ensemble series, the delays of the spikes in the target ensemble
series within -50 to +50 bin lags were calculated. Next, the
collective count of target spikes within a given lag was defined as
the value of ensemble cross-correlogram between FS and RS series. This
value was turned into a percentage for enabling the comparison across
subjects with different number of neurons, multiple scales with a
different number of aggregate of spikes, and different states with
different duration of events. This process was realized for all
scales.

\subsubsection*{Randomization}
Randomization was used to construct control for the ensemble cross-correlogram. We used four different systems of randomization to test for different within and between aspect of ensemble series. Any of the randomization protocols was realized 100 times. For each randomization category, the average of 100 random ensemble cross-correlogram was used as the control for verification of the observed patterns in the non-randomized cross-correlogram.
\begin{itemize}
\item \textit{Random permutation of ISI in the ensemble series.} After pooling all the FS and RS cells into their ensemble series, the ensemble ISI (inter-spike interval) was calculated. Then, for each of the two ensemble series, a random permutation of its ISI was followed by cumulative summation of ISI, resulting in the new temporal order of ensemble spikes. This procedure guarantees that the randomized ensemble series has the exact number of spikes and exact set of ISIs as of the original ensemble series, albeit with different temporal arrangement of spikes within a given ensemble series. 
\item \textit{Circular shift of spike ensemble.} In this type of randomization, the spikes were first pooled to create the ensemble FS and RS series. For each series, the ISI of the ensemble series was calculated. Then all the spikes in each series were shifted at once with a random value between the lower bound (1) and upper bound (maximum of the ISI in the ensemble series). In each randomization trial, it was made certain that the degree of the shift was not equal for the two FS and RS ensemble, guaranteeing that the temporal relation of the two series was never repeated. In contrast to the previous procedure (random permutation), this randomization kept the temporal of order of spikes within each ensemble series same as the non-randomized series. However, here the temporal relation of the two FS and RS ensembles was disrupted. 
\item \textit{Fixed-ISI circular shift of spikes.} Before aggregating the spikes into the ensemble series, the ISI of each unit's spike series was calculated. Then the spikes of the unit were shifted based on a random value drawn between the lower bound (1) and upper bound (maximum of the ISI of the that unit's spike series). Next, all the randomized units were aggregated to create the randomized ensemble series. This procedure guarantees that the resultant ensemble series is constructed from units with intact internal structure of their spike timing but with a disrupted between-unit timing.
\item \textit{Local jitter randomization of spikes.} Next we tested the effect of randomization based on the statistics of each individual neuron before their aggregation to the ensemble series. First, the ISI of each FS (or RS) unit was calculated. Then the pool of the ISI as well as the ensemble of FS and RS was created. Next, each spike in the ensemble was shifted according a random number which was generated as the standard deviation plus a randomized (between -1 and 1, not including 0) multiple of the mean of pooled ISI. If the drawn random value was negative, the spike was shifted to the left and if the random chosen value was positive, the shift was toward the right in the ensemble series.  This randomization, guarantees a tightly regulated data-driven local randomization based on the statistical properties of individual spikes.
\end{itemize}

\subsection*{Deviation from absolute symmetry} 
We used complementary methods to calculate the deviation from symmetry between excitatory and inhibitory activities. First, we estimated the data-derived axis of symmetry based on the weighted bisquare robust regression. Then the angle between this axis and the identity line was used to represent the degree of deviation from pure symmetry. In parallel, for each time-scale, and for a given state, the time series of ensemble spiking data were reshaped into a 3-dimensional surface where the dimensions were the fraction of excitation, the fraction of inhibition, and number of their occurrences. As the durations of different states (SWS, REM, Wakefulness) differ from each other, the joint probability of ensemble fractions were also normalized by the whole length of the state to provide comparable results for further quantifications; i.e., the result is a surface in the 3D space of the fraction of excitation, the fraction of inhibition, and their joint probabilities. Theses surfaces were then Z-scored and their major orientation axis was calculated. Then the mid-point of the iso-surfaces along the major orientation axis was defined. Using orthogonal regression, a plane was fit to these point along the major orientation axis. This plane, is the plane of approximate symmetry of the data and divides the surface into two halves. In case of absolute balance at a given scale, the plane of symmetry of data would coincide with the symmetry plane of the 3D space. Deviations from perfect balance was calculated using the dihedral angle between the symmetry plane of data and symmetry plane of the 3D space. The results of the dihedral rotation was similar to the angle between axis of symmetry and the weighted least square regression (using robust bi-square fit) of the data in the 2D rendering of excitation fraction and inhibition fraction.

\subsection*{Computational model}

Network simulations were done using networks of excitatory and
inhibitory spiking (integrate-and-fire type) neurons with sparse
random connectivity (2000 excitatory and 2000 inhibitory neurons with
5$\%$ connection probability), and with conductance-based (COBA)
synaptic interactions (See Fig.\ref{fig:COBAbalanceNew}A).  Such COBA
networks were shown to display self-sustained asynchronous irregular
(AI) balanced states (see \cite{Vogels2005} for
details of the parameters and see \cite{Brette2007} for codes). The
network activity was entirely self-sustained (no added noise), after a
kickoff random simulation to initiate the AI state. For creating an externally-driven network, we used the same method with 10-times reduced collateral synaptic weights, but receiving a noisy external input.

\subsection*{Acknowledgments}
The research was funded by Wyss Institute for Biologically-Inspired Engineering at Harvard University, Centre National de la Recherche Scientifique (CNRS, France), Agence Nationale de la Recherche (ANR, France), European Community Future and Emerging Technologies program (BrainScales FP7-269921; The Human Brain Project FP7-604102) and National Institutes of Health (NIH grants 5R01NS062092, R01EB009282 and R01NS045853).
  
\subsection*{Author contributions statement}
N.D and A.D conceived the study. S.C, E.H and N.H conducted the experiments. N.D analyzed the experimental data with contributions from A.P, B.T and M.LVQ. A.D and N.D performed the simulation and its analysis. N.D and A.D interpreted the results and wrote the paper with contributions from all authors.

 
 \nocite{*}

\newpage
\beginsupplement
 \setcounter{page}{1}
 \pagenumbering{roman}
\begin{center}
\textbf{Supplementary Material}
\end{center}

\begin{figure}[h!]
\begin{center}
\includegraphics[width=0.9\columnwidth]{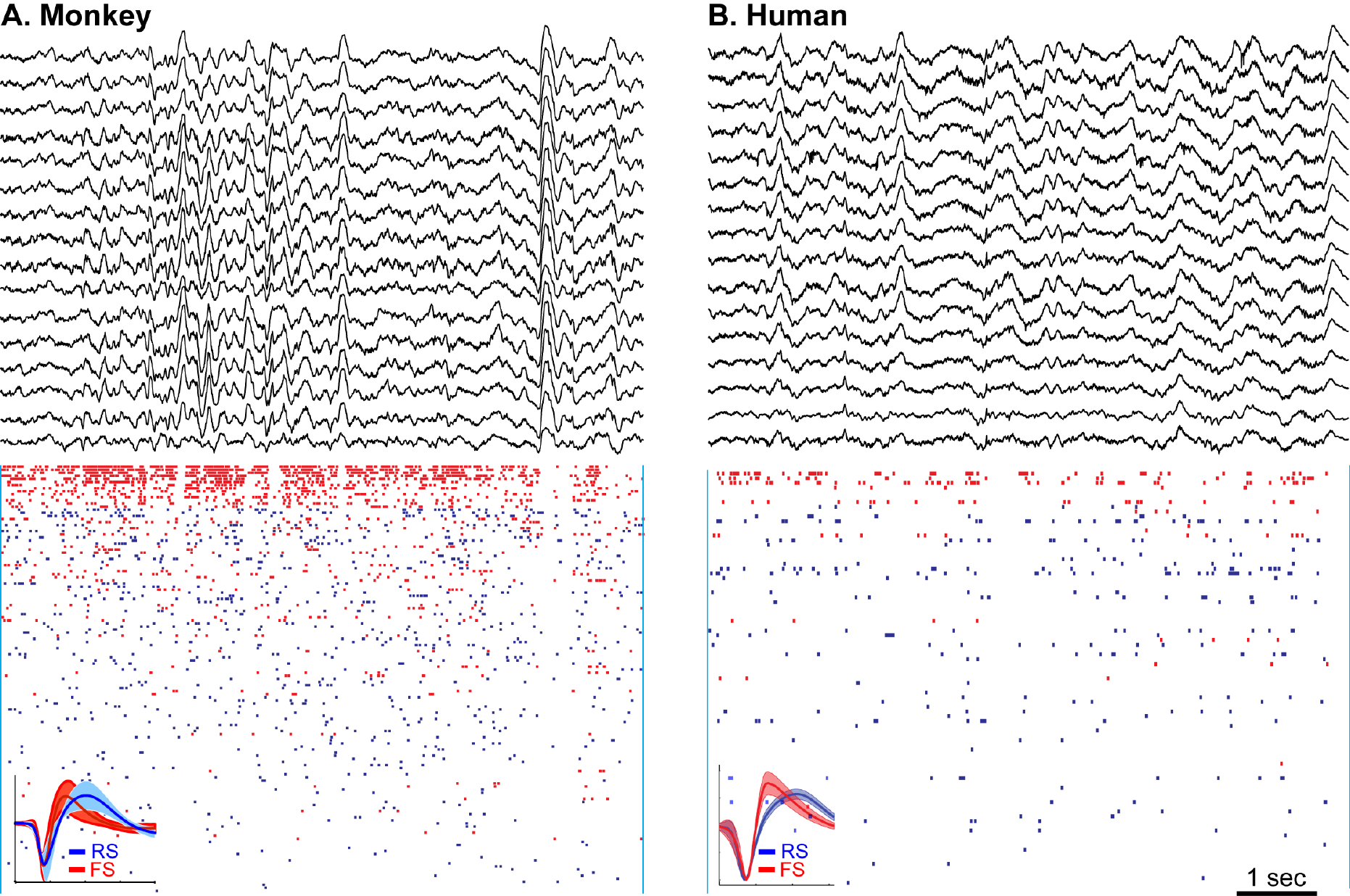}
\caption{\label{fig:RawTrace} Sample recordings from a UTAH
  multielectrode array in Monkey (A) PMD and Human (B) Temporal
  cortices during 8 sec of SWS (Slow-wave sleep). In each panel, the
  upper section depicts LFP (local field potentials) from different
  locations of the multielectrode array. Lower sections show the
  corresponding Excitatory (blue) and Inhibitory (red) cells. Insets
  show the spike-waveform that was used to categorize the units into
  two inhibitory and excitatory cell populations.%
}
\end{center}
\end{figure}

\begin{figure}[h!]
\begin{center}
\includegraphics[width=0.9\columnwidth]{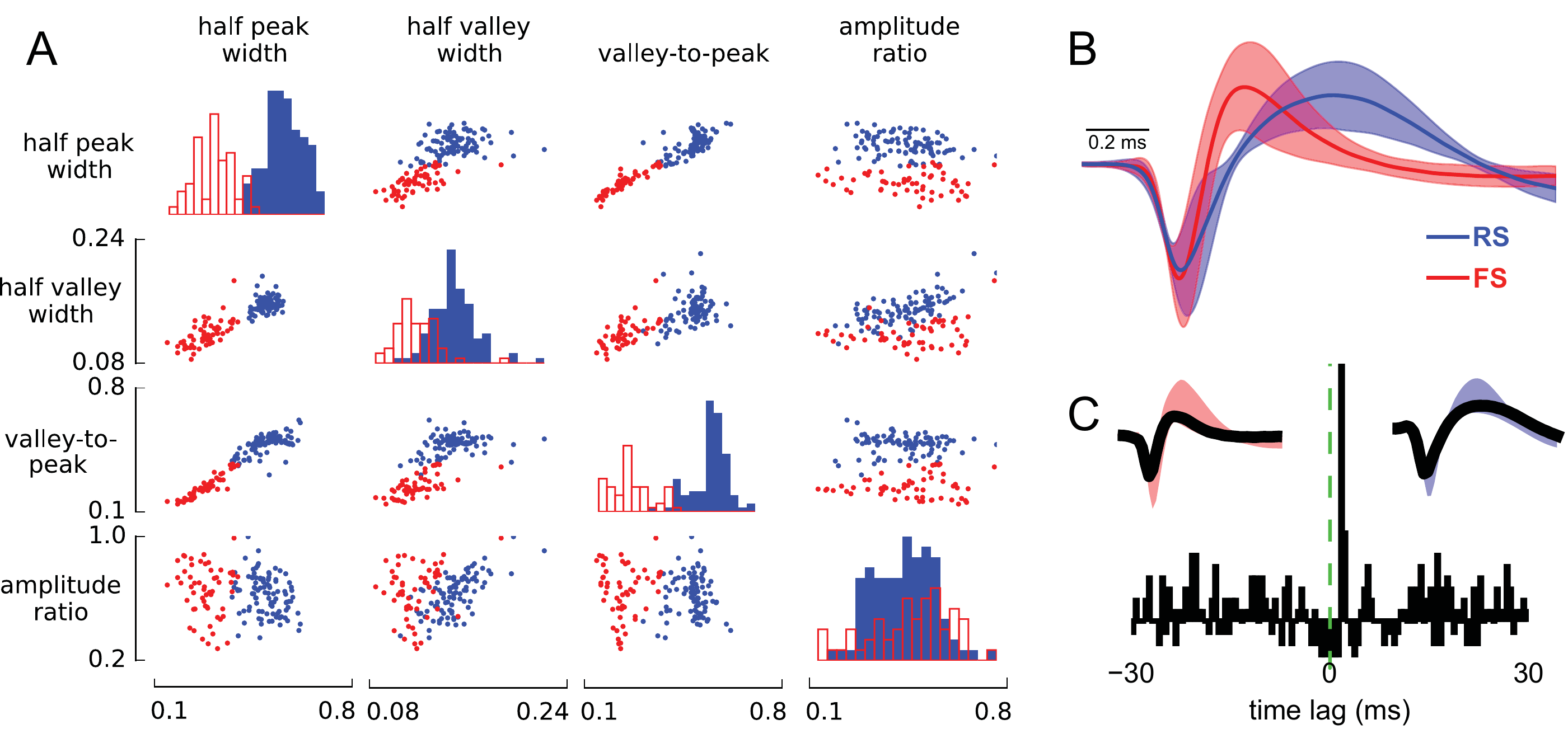}
\caption{\label{fig:FSRSdetails} Separation between the putative regular spiking and fast spiking neurons in monkey. (A) Clustering based on 4 features of the spike waveform (half-peak width, half-valley width, valley-to-peak amplitude and valley-to-peak amplitude ratio) and we classified the spike waveforms in this 4-dimensional feature space into two classes (putative regular spiking, RS, and fast spiking, FS). The data points and histograms corresponding to these two types are shown in different colors (RS - blue, FS - red). The off-diagonal panels represent scatter plots of between each pair of features (labels at the top and left edges of the figure, each dot corresponds to a single neuron). The bar plots on the diagonal represent the histograms of spike-waveform features for each neuron type separately.  (B) Average spike waveforms of the RS and FS type. (C) An example cross-correlogram (black bars, bin size 0.5 ms) between a pair of putative fast-spiking and regular-spiking neurons (spike waveforms shown in insets). The positive lags represent spikes of the FS neuron arriving after the spike of the RS neuron. The prominent peak at 2 ms corresponds to a presumed mono-synpatic connection from the RS neuron to the FS neuron, confirming functionally the separation based on electrophysiological features (spike waveform).%
}
\end{center}
\end{figure}

\begin{figure}[h!]
\begin{center}
\includegraphics[width=0.6\columnwidth]{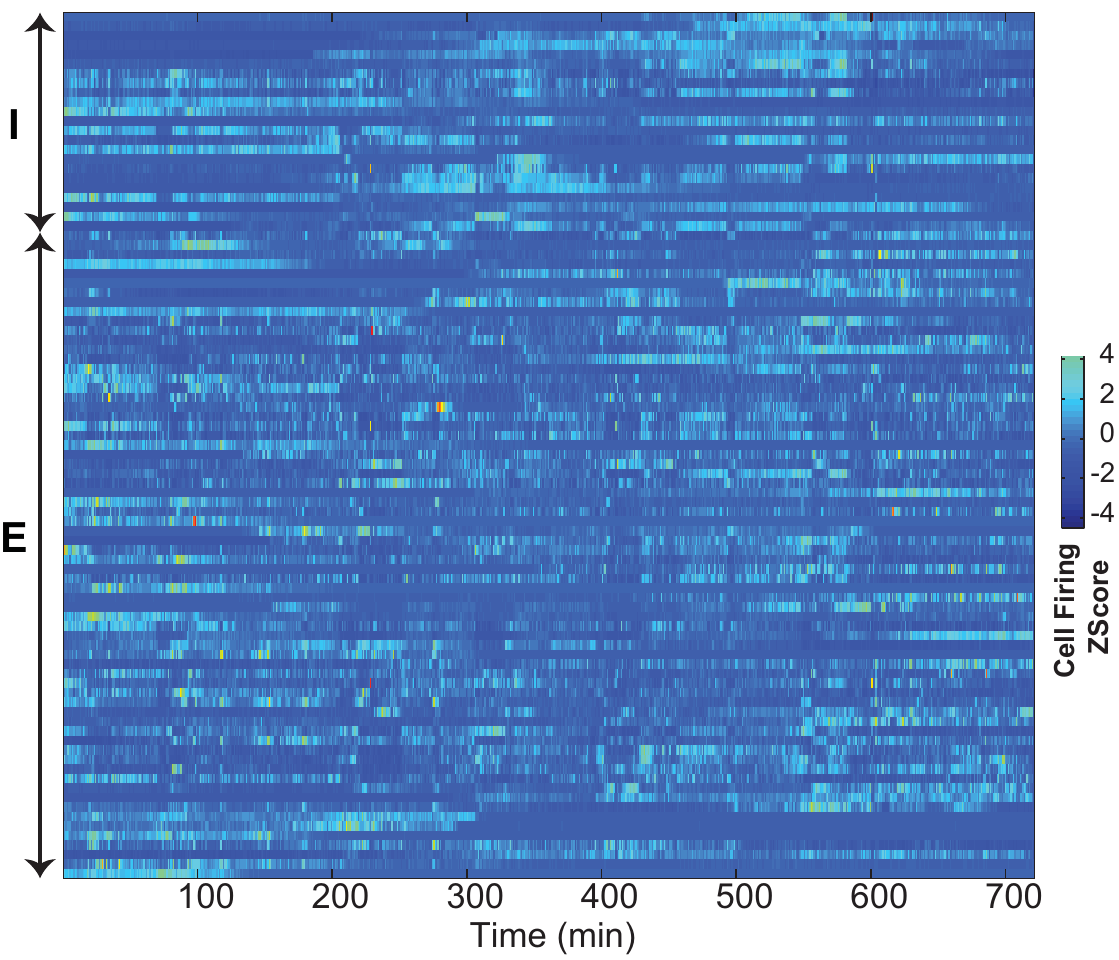}
\caption{\label{fig:firingVariability} Firing variability during a 12 hour recording for I (N=23) and E (N=68) cells in human patient 1 (each cell firing cells is z-scored through the 12 hour recordings).%
}
\end{center}
\end{figure}

\begin{figure}[h!]
\begin{center}
\includegraphics[width=0.7\columnwidth]{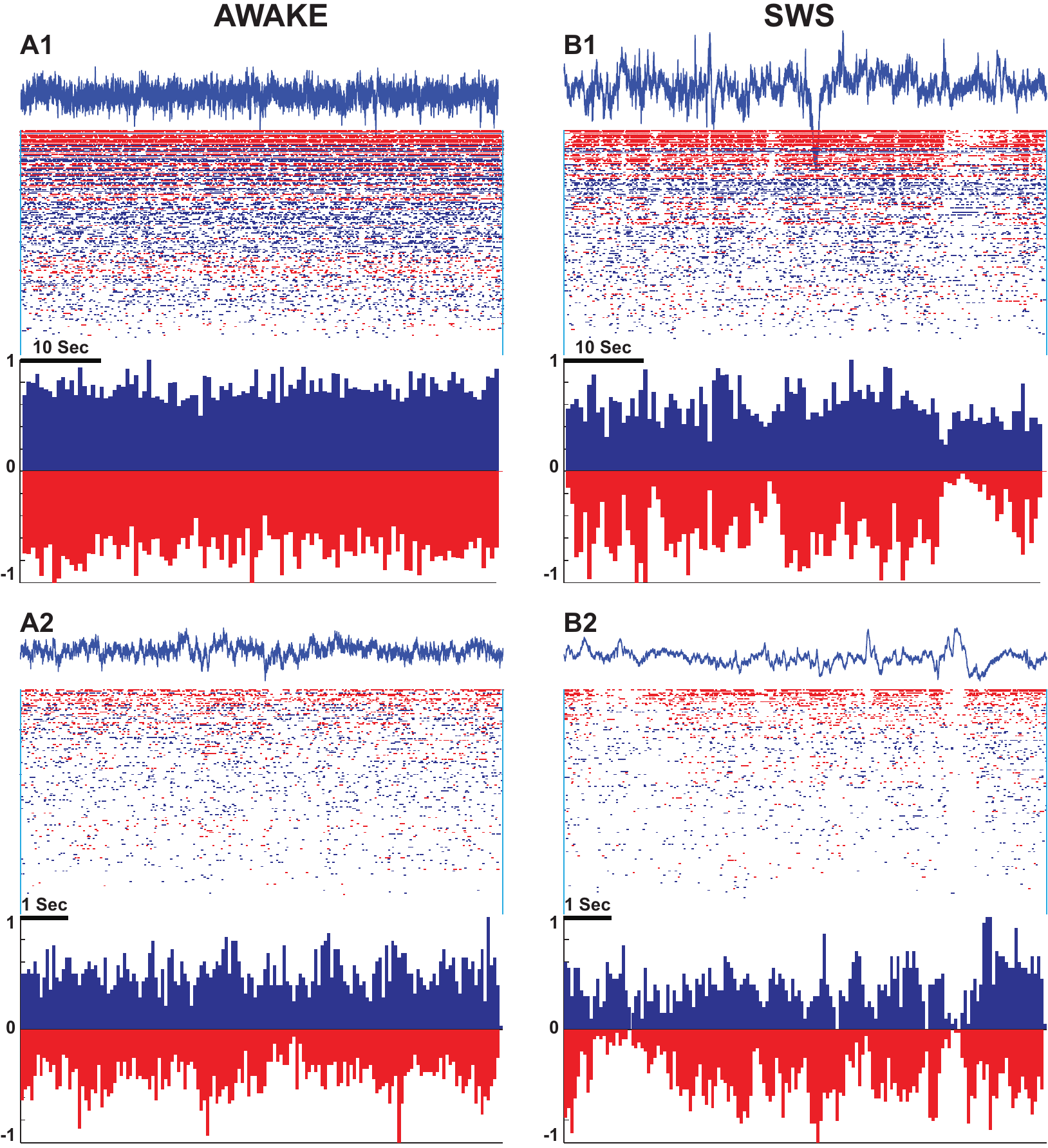}
\caption{\label{fig:MonkeyStateTrace} Sample recordings for AWAKE
  (left) and SWS (right) in monkey. A1 and B1 show 60 seconds windows;
  A2 and B2 show a 10 second window of the same state. In the rasters,
  putative inhibitory neurons (FS cells) and putative excitatory
  neurons (RS) are depicted in red and blue, respectively. In each
  panel, a sample LFP accompanies the spiking activity. Neurons are
  sorted based on their firing rate within the 60 sec epochs, in a
  descending order. Histograms show the overall excitatory activity
  normalized to the maximum of firing rate (within FS or RS category)
  in the shown example.%
}
\end{center}
\end{figure}

\begin{figure}[h!]
\begin{center}
\includegraphics[width=0.5599999999999999\columnwidth]{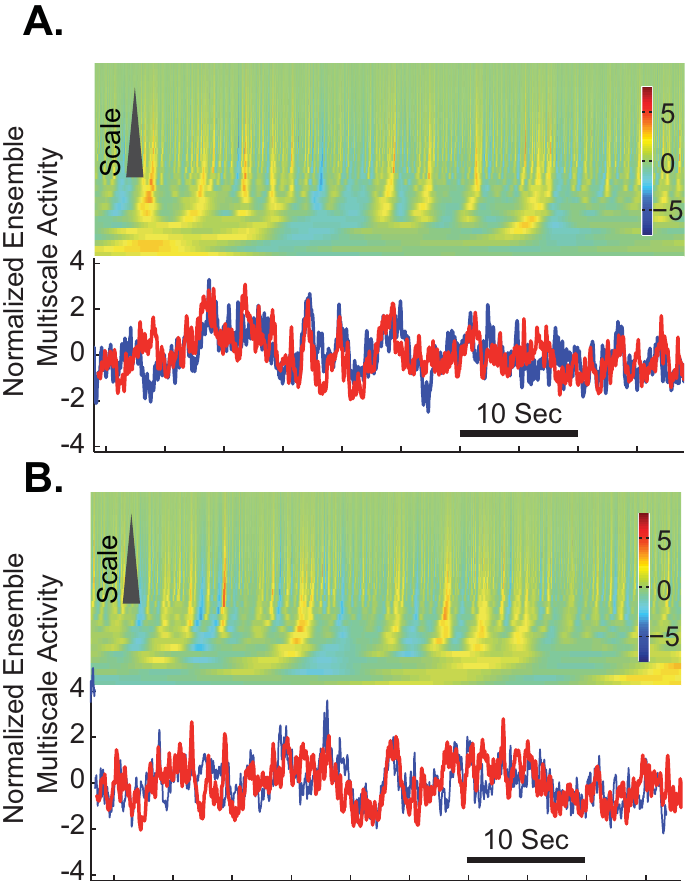}
\caption{\label{fig:monkeyMultiscaleTrace} Multiscale features of
  excitation and inhibition balance in two sample recordings from the
  monkey. As in Fig.\ref{fig:rawMultiscaleTrace}, each row in the
  heatmap shows the normalized (Z-score) difference of ensemble
  excitation and inhibition for a given scale. The scales are defined
  as in Fig.\ref{fig:rawMultiscaleTrace}, increasing from the top to
  bottom (from fine-grain to coarse-grain). The color saturation
  towards red signifies instantaneous dominance of inhibition. Blue
  saturation shows instantaneous dominance of excitation, while green
  shows tight match between normalized ensemble excitation and
  inhibition. Line traces show the Z-scored addition of normalized
  excitatory (blue) and inhibitory (red) ensembles across the scales.%
}
\end{center}
\end{figure}

\begin{figure}[h!]
\begin{center}
\includegraphics[width=0.9\columnwidth]{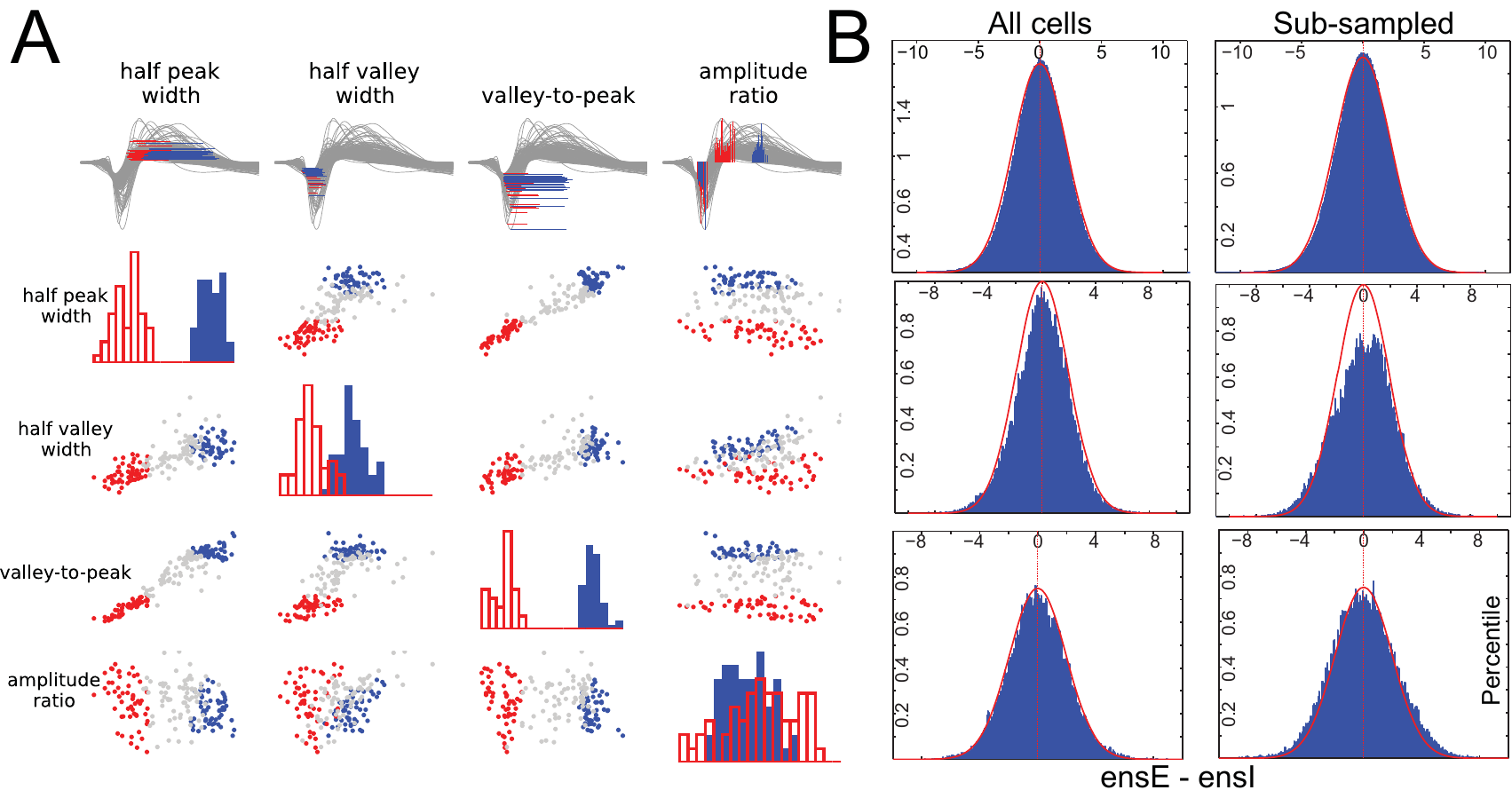}
\caption{\label{fig:FSRSsubsample} A. Scatter plots of pairs of spike-waveform features (off the diagonal) and histograms of individual features (on diagonal). The waveforms were clustered into RS (blue) and FS (red) types. Compared to Supplementary Fig.\ref{fig:FSRSdetails}, we only take the waveforms with the 30\% highest confidence of the clustering (the confidence being defined as the positive or negative projection on the vector connecting cluster center).  B. Histogram of multiscale ensemble E and ensemble I difference at at given time t shows that balance of E and I is not affected by taking only the most clear FS and RS cells. The red curve is the scaled pdf of normal distribution calculated from the original data (left). Left column, original data. Right column, sub-set of neurons from 30\% both ends of the FS-RS classification spectrum. Top row, whole data sets. Middle and bottom row, data in the segment from supplementary Fig.\ref{fig:monkeyMultiscaleTrace}.A and B, respectively.%
}
\end{center}
\end{figure}

\begin{figure}[h!]
\begin{center}
\includegraphics[width=0.84\columnwidth]{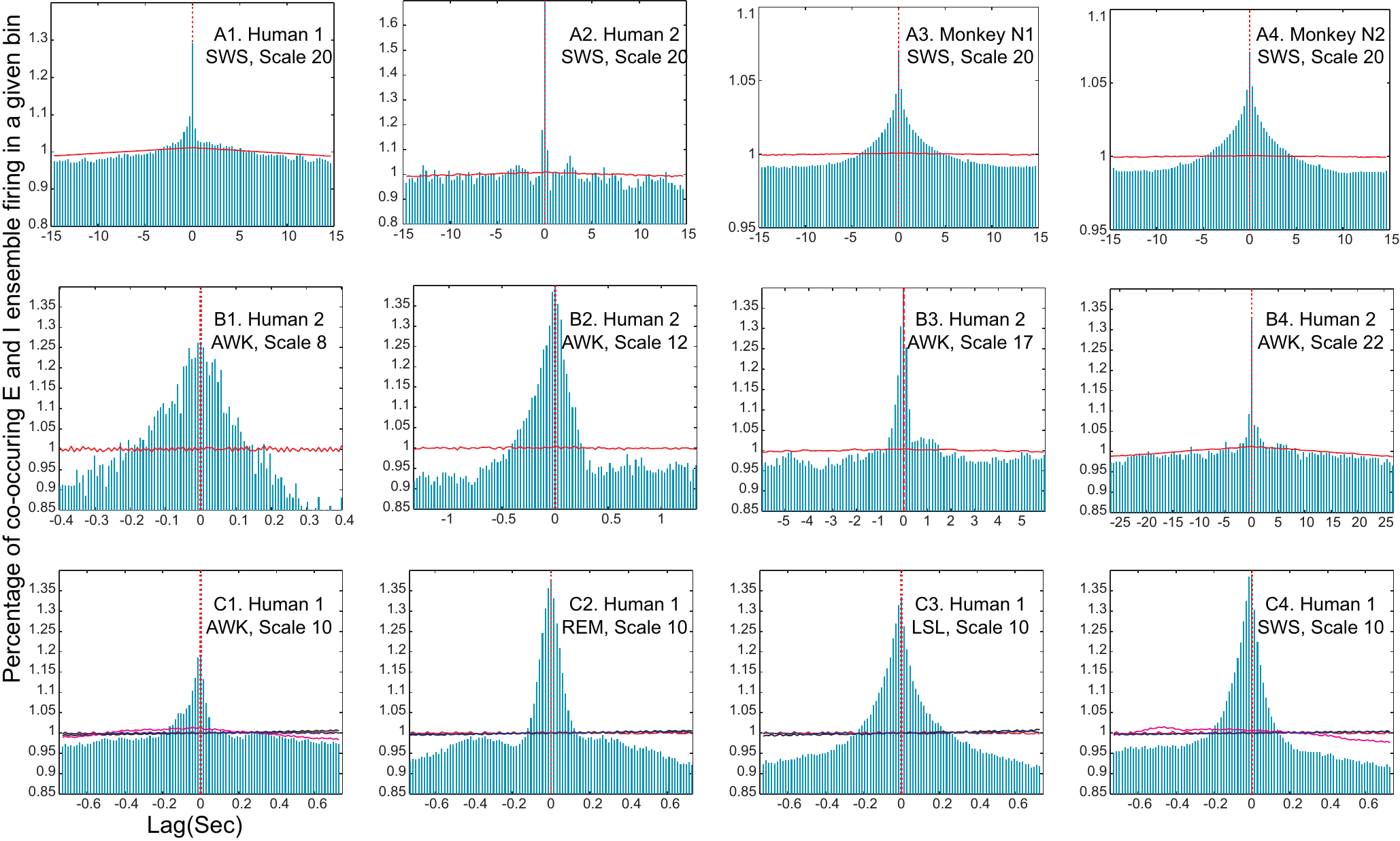}
\caption{\label{fig:ensxcorr}. Excitation and inhibition are
  correlated over multiple scales.  In each panel, the
  cross-correlogram is shown as the histogram of delays of the spikes
  in the ensemble target series (inhibitory) with respect to the
  spikes of the reference series (excitatory). The vertical dashed
  line shows the lag zero, the horizontal line shows the average
  ensemble cross-correlogram of the Monte Carlo randomized process. In
  each histogram, the count of delays is turned into percentage
  (y-axis) for comparative reliability across different subjects (with
  different number of cells), different scales (different bin sizes)
  and different states (different length of the event). Note that in
  all panels lags -50 to +50 (bins) are shown. However, the span of
  time (x-axis, in sec) depends on the bin size of the evaluated
  timescale. A1 to A4, Ensemble cross-correlograms during slow-wave
  sleep across two different humans and two different nights of
  recording from the monkey are shown for a sample timescale. The
  shown randomized control (red) is the average of 100 realization of
  random permutation of the ensembles (see methods). B1 to B4,
  Ensemble cross-correlogram during wakefulness for a given subject
  across four different scales. Note that in each histogram of delays,
  the same number of lags (-50 to +50) are tested.  The randomized
  control (red) is the average of 100 rounds of realization of random
  local jitter (see methods for details). C1 to C4, Ensemble
  cross-correlogram of different states in another human subject for
  an example scale. All four randomized controls (horizontal lines)
  show similar outcomes. The randomized controls show that these four
  different randomization procedures yield highly reliable dispersion
  of events in the ensemble series such that the ensemble
  cross-correlogram no longer shows any temporal interdependency
  between the ensemble excitatory and inhibitory series.%
}
\end{center}
\end{figure}

\begin{figure}[h!]
\begin{center}
\includegraphics[width=0.7\columnwidth]{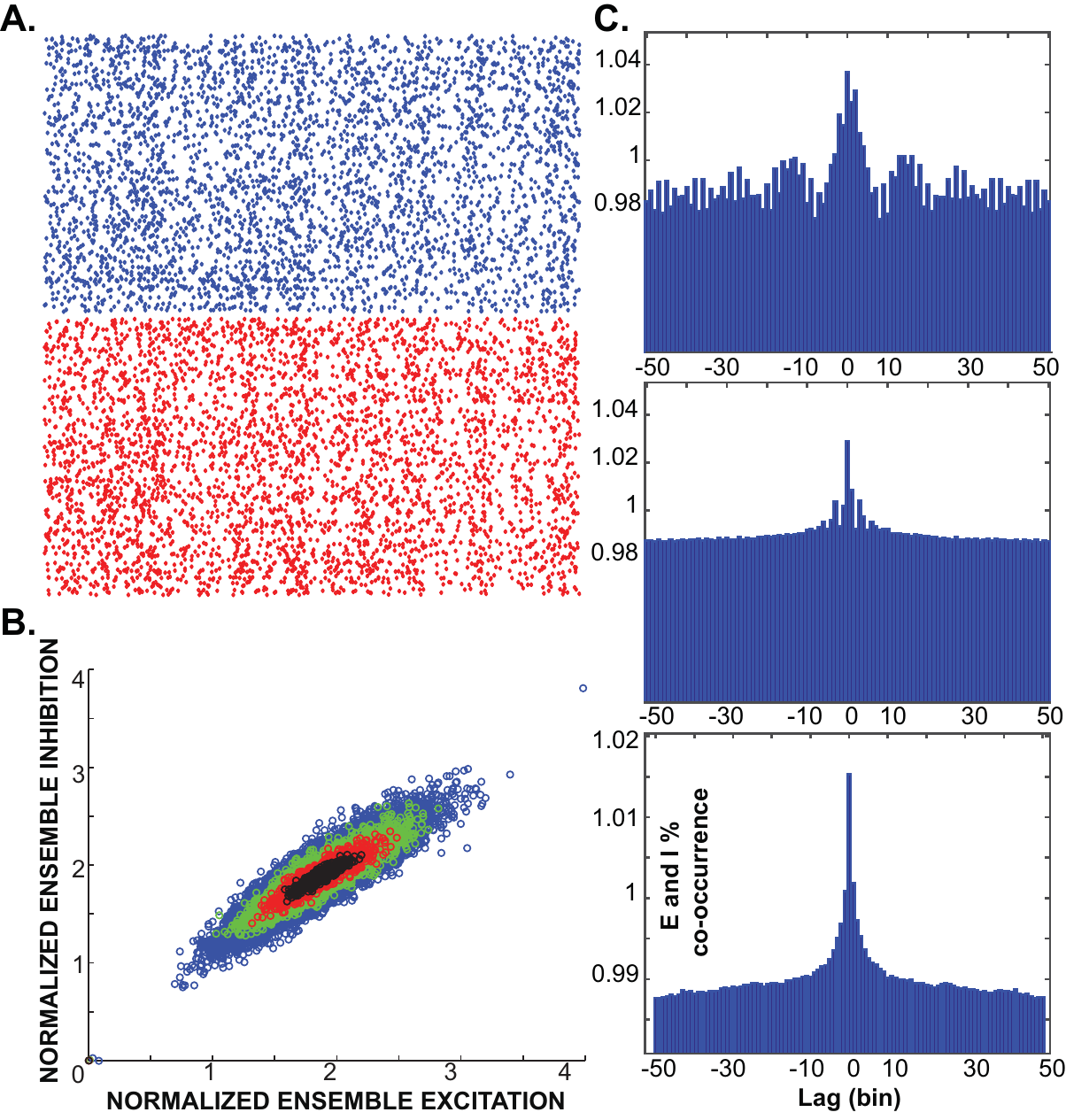}
\caption{\label{fig:COBAbalanceNew} Balanced activity states in
    model networks of excitatory and inhibitory spiking neurons. A.
  Raster of 2000 excitatory (blue) and 2000 inhibitory (red) in a COBA
  (conductance-based) model, showing AI (asynchronous irregular)
  state for 100 sec. Panel B shows the scatter plot of normalized ensemble
  excitation and inhibition for times $t_{1}$ to $t_{n}$, with n
  representing the length of the time series at a given scale. Blue,
  red, green and black represent sample scales (comparable to
    scales used in the data; see Fig.\ref{fig:rawMultiscaleTrace}).
  Similar to the experimental data, the paired E-I data scatter along
  the diagonal line, representing an overall balance across multiple
  scales, with moment to moment dominance of E or I, while the degree
  of scattering along the diagonal shrinks with coarse-graining. C. Ensemble cross correlation score for three sample scales show preserved balance at central lag.%
}
\end{center}
\end{figure}

\begin{figure}[h!]
\begin{center}
\includegraphics[width=0.8\columnwidth]{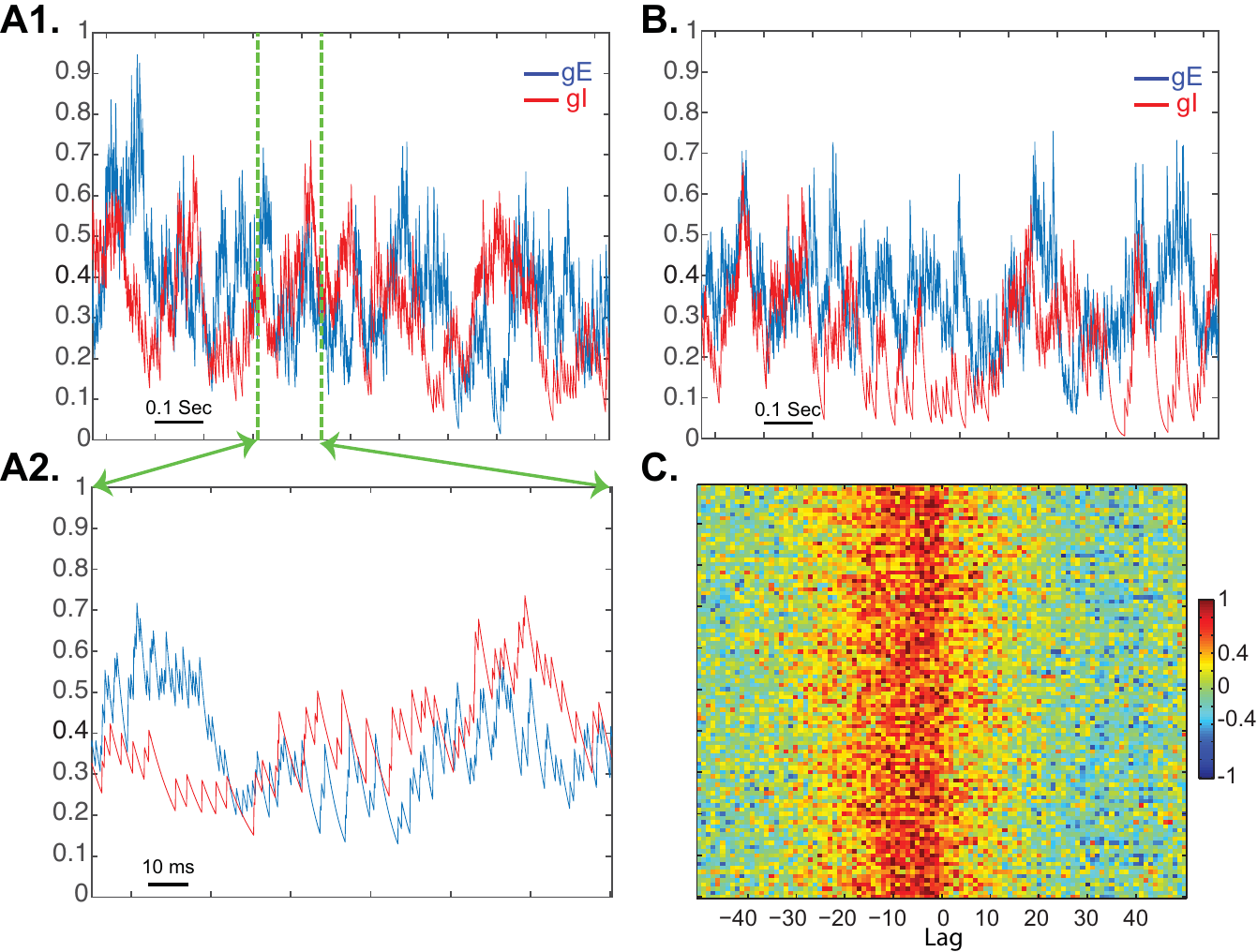}
\caption{\label{fig:conductance} A1,B. Normalized $g_{E}$ (excitatory conductance, blue) and $g_{I}$ (inhibitory conductance, red) for a sample segments for two example E cells. A2, zoom in the marked region of A1. C. Correlation of $g_{E}$ and $g_{I}$ in 100 sample cells from the model. Each row shows normalized cross correlation between $g_{E}$ and $g_{I}$ for 50 lags (each lag = 1ms). Cells show variable conductance correlation maximum lag, with $g_{I}$ lagging behind $g_{E}$ on average. D. The cross-correlation between two exponential kernels provides similar characteristics to the $g_{E}$:$g_{I}$ correlation. Inset, exponential kernels constructed with decay time based on average $g_{E}$ and $g_{I}$ rise time (across all cells) and delayed based on the average $g_{E}$:$g_{I}$ conductance correlation.%
}
\end{center}
\end{figure}

\begin{figure}[h!]
\begin{center}
\includegraphics[width=0.9\columnwidth]{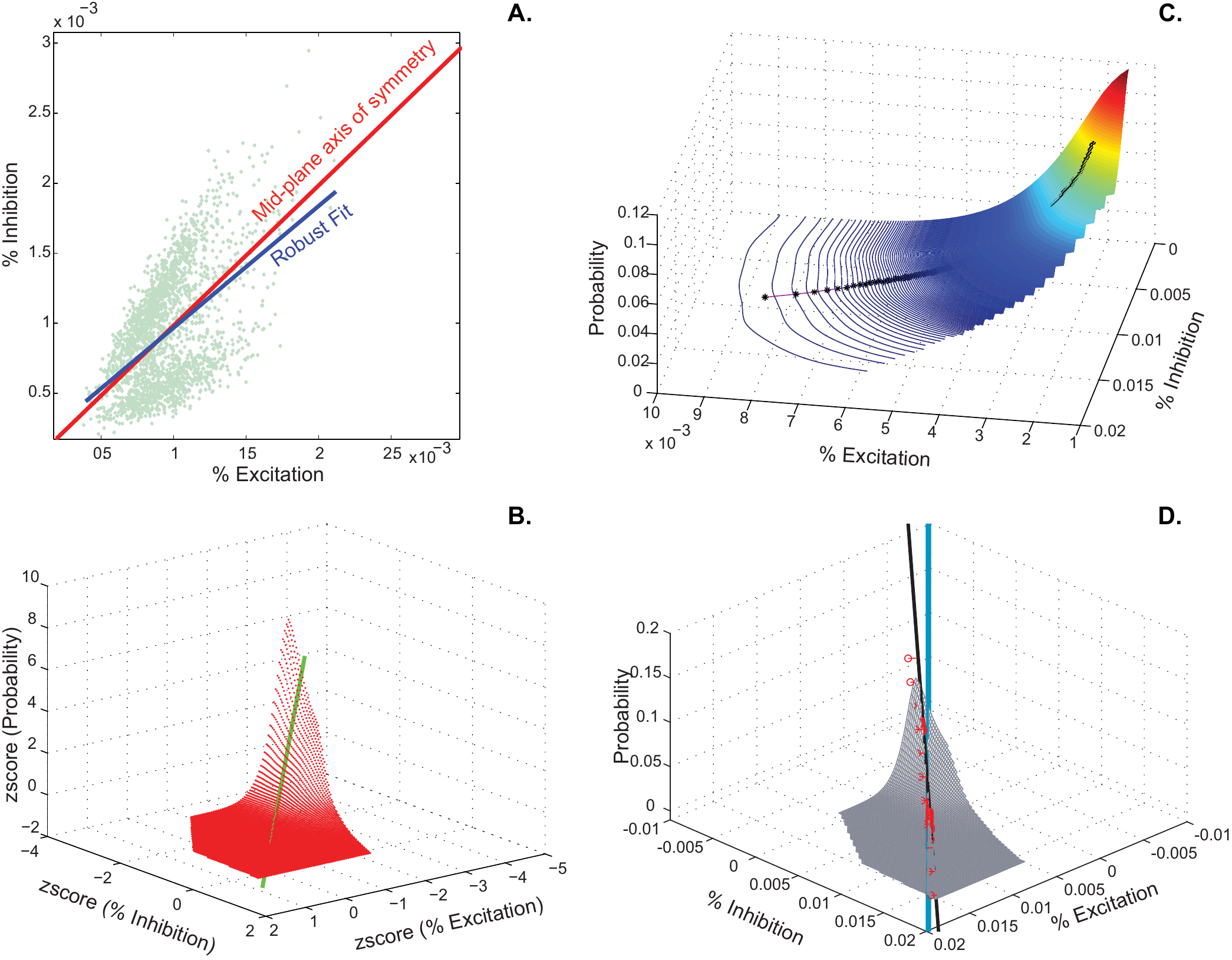}
\caption{\label{fig:jointProbEI} Panel A shows estimation of deviation
  from balance, between ensemble excitation and inhibition for a
  sample scale of SWS in a human subject, using robust bisquare
  regression. The fit (blue line) to the green cloud (data) shows the
  axis of symmetry of the data. Its deviation from the symmetry axis
  of the plane (in red) shows the degree of balance deviation. Panels
  B to D each show a different method for estimating the deviation
  from perfect symmetry. Panel B, shows the major orientation axis of
  the Z-scored data. Panel C, shows the distribution of E-I ensemble
  fraction pairs for a sample scale during SWS. The black lines are
  the centroids of the iso-surfaces. Panel D, combining these info,
  one can find the mid-plane of the data (shown in black) and find its
  tetrahedral angle with the plane of absolute symmetry (shown in
  cyan; see also the distribution of such angles in
    Fig.\ref{fig:BalanceDeviation}).%
}
\end{center}
\end{figure}

\begin{figure}[h!]
\begin{center}
\includegraphics[width=0.8\columnwidth]{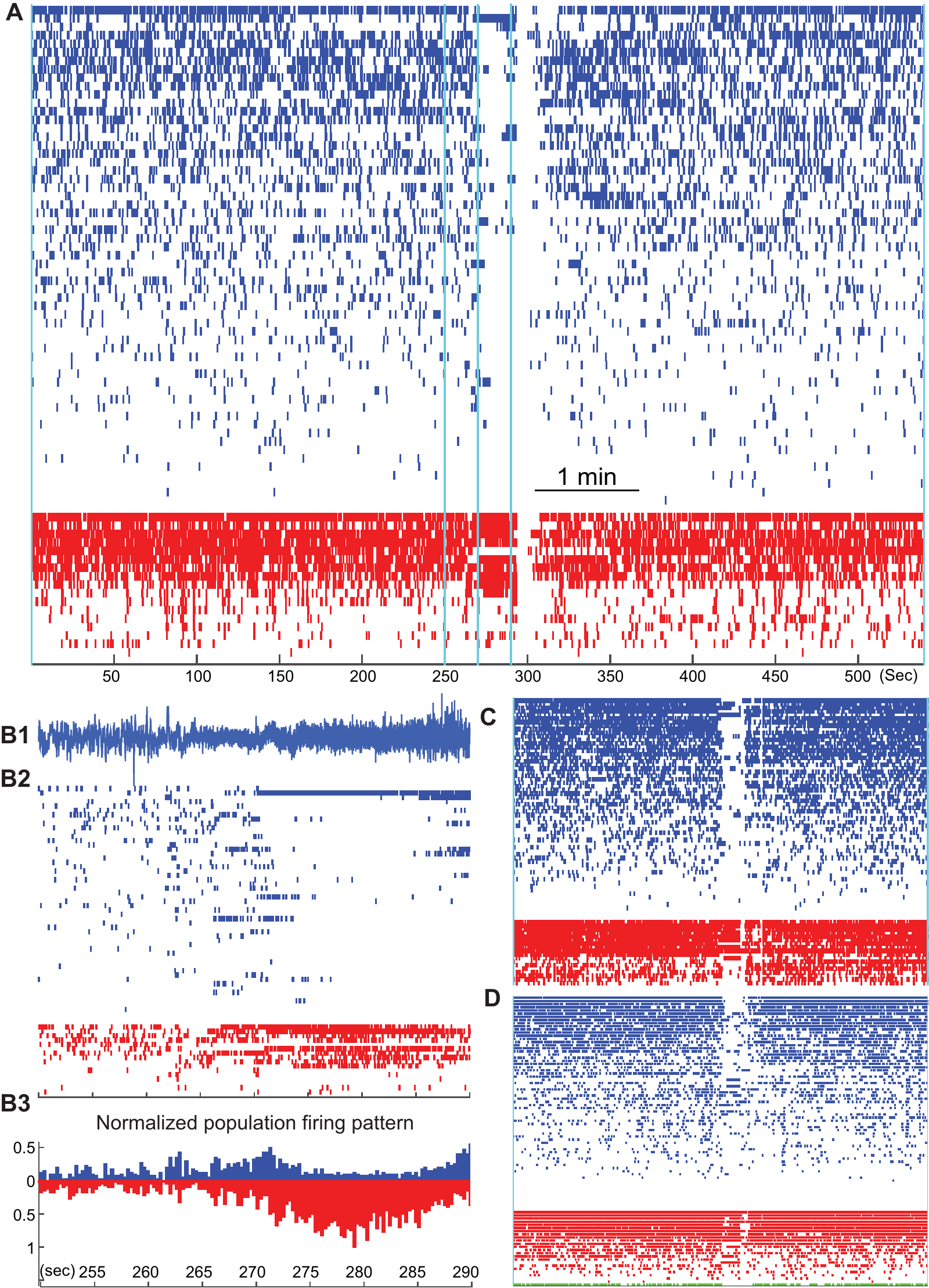}
\caption{\label{fig:seizureXtra} Misbalance in more example seizure
  recordings in human. Panel A shows a 9 minute recording.  Panels in
  B are the zoomed in version (the middle 40 seconds) of the same
  epoch (shown with the vertical lines in A). RS cells are in blue and
  ranked based on their firing rate within this epoch.  Red cells show
  FS cells and are ordered according to their class firing rate. B1,
  LFP activity in the zoomed period, corresponding to B2 raster of FS
  and RS cells. B3, Normalized mirrored histogram showing where the
  misbalance occurs. C, D. Two additional seizures similar to panel A
  (Note on panel D: in this patient, one unit was not categorized as
  either FS or RS, shown in green). Seizure happens around the
  mid-point and is visually distinct from the rest of the recording.%
}
\end{center}
\end{figure}

\begin{figure}[h!]
\begin{center}
\includegraphics[width=0.8\columnwidth]{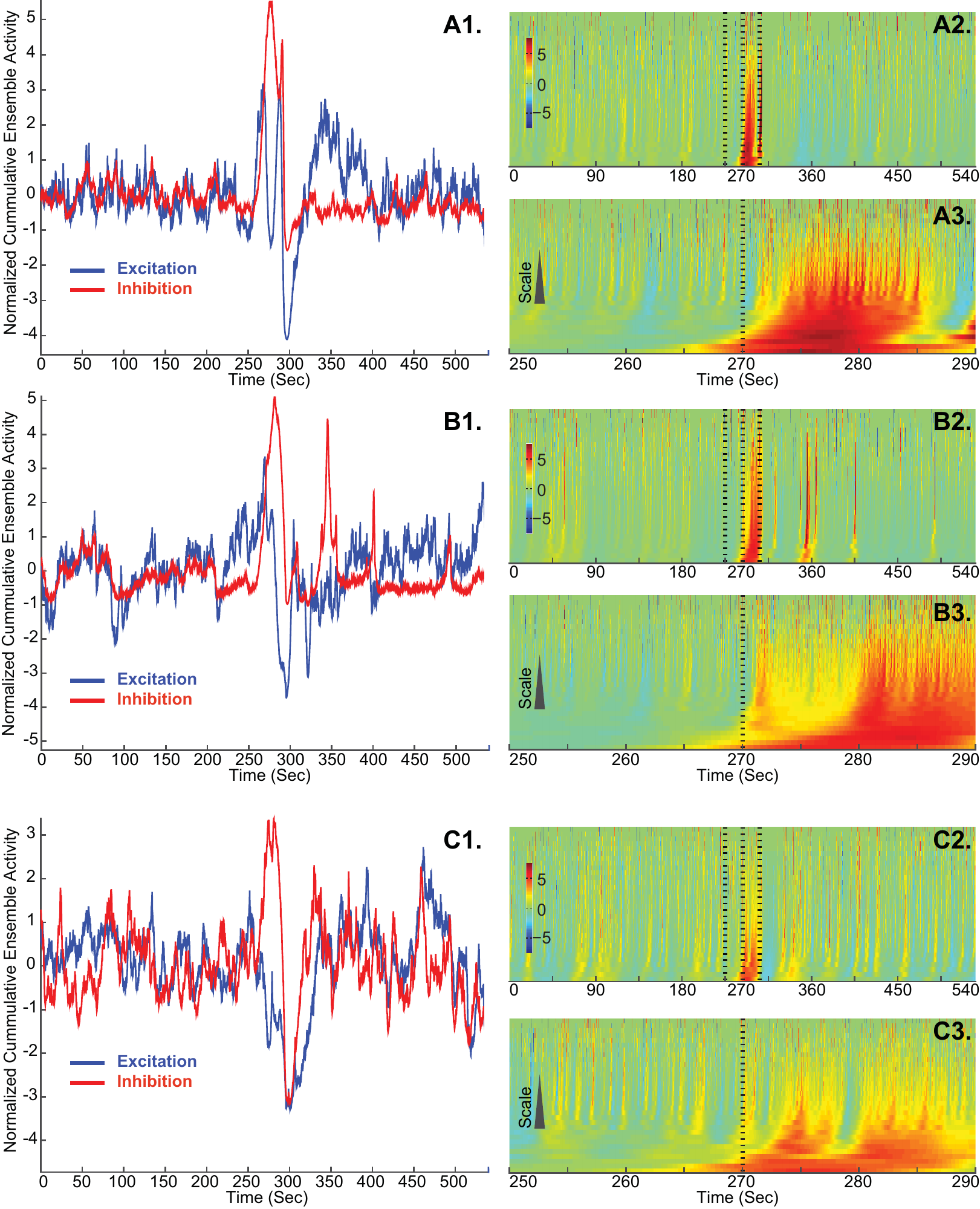}
\caption{\label{fig:SeizureMultiscale}  Break down of E/I balance in different seizure episodes. A1, B1 $\&$ C1. Multiscale features of balance breakdown during seizure. Blue and red traces show the normalized cumulative activity of ensemble excitation and inhibition across multiple scales (similar to Fig.\ref{fig:rawMultiscaleTrace} bottom panels). In the shown examples (as well as in Fig.\ref{fig:SeizureMisbalance}), electrographic seizure starts around 270 sec. In all cases, ensemble excitation and inhibition follow the same multiscale trend. At the time of seizure, the two ensembles go through major fluctuations, and disentangle. In C1, return to multiscale balance trend happens fairly shortly. In A1, the system returns to balance a bit later (around second 400) and in B1, the system shows prolonged disturbed balance in the examined period shown here. Panels A2:3, B2:3 and C2:3 show the heatmap of the normalized ensemble excitatory and inhibitory differences, corresponding to the 9 minute recording and the middle 40 seconds zoom in (similar to Fig.\ref{fig:SeizureMisbalance}C).%
}
\end{center}
\end{figure}

\end{document}